\documentclass{ws-ijbc}
\usepackage{ws-rotating}     
\usepackage{graphicx}
\usepackage{epstopdf}

\newcommand{\be}{\begin{equation}}
\newcommand{\ee}{\end{equation}}
\newcommand{\Dlt}{\Delta}
\newcommand{\dlt}{\delta}
\newcommand{\prt}{\partial}

\newcommand{\bt}{\beta}

\newcommand{\ep}{\varepsilon}
\newcommand{\al}{\alpha}
\newcommand{\ra}{\rightarrow}
\newcommand{\sgm}{\sigma}

\begin{document}

\catchline{}{}{}{}{} 

\markboth{V.I. Yukalov, E.P. Yukalova, D. Sornette}{Population dynamics with 
nonlinear delayed carrying capacity}

\title{POPULATION DYNAMICS WITH NONLINEAR DELAYED CARRYING CAPACITY}

\author{V.I. YUKALOV}
\address{Department of Management, Technology and Economics, \\
ETH Z\"urich, Swiss Federal Institute of Technology,
Z\"urich CH-8092, Switzerland\\ and \\
Bogolubov Laboratory of Theoretical Physics, \\ 
Joint Institute for Nuclear Research, Dubna 141980, Russia \\
yukalov@theor.jinr.ru}

\author{E.P. YUKALOVA}
\address{Department of Management, Technology and Economics, \\
ETH Z\"urich, Swiss Federal Institute of Technology,
Z\"urich CH-8092, Switzerland\\ and \\
Laboratory of Information Technologies, \\
Joint Institute for Nuclear Research, Dubna 141980, Russia \\
lyukalov@ethz.ch}

\author{D. SORNETTE}
\address{Department of Management, Technology and Economics, \\
ETH Z\"urich, Swiss Federal Institute of Technology,
Z\"urich CH-8092, Switzerland \\ and \\
Swiss Finance Institute, c/o University of Geneva, \\
40 blvd. Du Pont d'Arve, CH 1211 Geneva 4, Switzerland \\
dsornette@ethz.ch}

\maketitle

\begin{abstract}
We consider a class of  evolution equations describing population dynamics
in the presence of a carrying capacity depending  on the population with 
delay. In an earlier work, we presented an exhaustive classification of the
logistic equation where the carrying capacity is linearly dependent on the 
population with a time delay, which we refer to as the
``linear delayed carrying capacity'' model. Here, we generalize it to the 
case of a nonlinear delayed carrying capacity. The nonlinear functional form 
of the carrying capacity characterizes the delayed feedback of the evolving 
population on the capacity of their surrounding by either creating additional 
means for survival or destroying the available resources. The previously 
studied linear  approximation for the capacity assumed weak feedback, while 
the nonlinear form is applicable to arbitrarily strong feedback. The 
nonlinearity essentially changes the behavior of solutions to the evolution 
equation, as compared to the linear case. All admissible dynamical regimes are 
analyzed, which can be of the following types: punctuated unbounded growth, 
punctuated increase or punctuated degradation to a stationary state, convergence 
to a stationary state with sharp reversals of plateaus, oscillatory attenuation,
everlasting fluctuations, everlasting up-down plateau reversals, and divergence 
in finite time. The theorem is proved that, for the case characterizing the 
evolution under gain and competition, solutions are always bounded, if the 
feedback is destructive. We find that even a small noise level profoundly 
affects the position of the finite-time singularities. Finally, we 
demonstrate the feasibility of predicting the critical time of solutions 
having finite-time singularities from the knowledge of a simple quadratic 
approximation of the early time dynamics.
\end{abstract}



\section{Introduction}

There exists a number of evolution equations characterizing population dynamics,
which have been applied to numerous concrete systems, ranging from populations 
of human and biological species to the development of firms and banks (see review
articles \cite{Kapitza_1,Hern_2,Korotayev_3,Yukalov_14}). The mathematical 
structure of these equations usually represents some generalizations of the 
logistic equation. Such equations can be classified into three main classes, 
depending on the nature of the dynamics of the carrying capacity. The first 
class, independently of whether the growth rate is a nonlinear function without 
or with delay of the population, assumes that the carrying capacity is a constant 
quantity given once for all that describes the total resources available to the 
population, in agreement with the initial understanding of the carrying capacity
(e.g., \cite{Haberl_5,Varfolomeyev_6,Hui_7,Gabriel_8,Berezowski_9,Arino_10}).
The second class allows the carrying capacity to change as a function of time, but
for exogenous reasons, either by explicitly prescribing its evolution or by specifying
its own independent dynamics for instance also given by a logistic equation
\cite{Dolgonosov_11,Pongvuthithum_12}.

The third class of equations interprets the carrying capacity as a functional 
of the population itself, implying that the population does influence the carrying
capacity, either by producing additional means for survival or by destroying the 
available resources \cite{Yukalov_13,Yukalov_14,Yukalov_4}. This feedback makes 
it possible to describe the regime of punctuated evolution, which is often observed 
in a variety of biological, social, economic, and financial systems 
\cite{Yukalov_13,Yukalov_14}.

In our previous articles \cite{Yukalov_13,Yukalov_14}, the carrying capacity was
approximated by a linear dependence on the population variable, which, strictly
speaking, assumes that the population influence should be smaller than the
initial given capacity. In the present paper, we generalize the approach by 
accepting a nonlinear carrying capacity that allows us to consider a population 
influence of arbitrary strength. Moreover, the population variable enters the 
carrying capacity with a time delay, since to either create or destroy resources 
requires time.

The outline of the present paper is as follows. In Sec. 2, we explain the 
deficiency that is typical of the linear capacity specification and suggest the 
way of generalizing its form to a nonlinear expression, by the use of the theory 
of self-similar approximations. The existence and general conditions for the 
stability of evolutionary stationary states are formulated in Sec. 3. The 
temporal behavior of solutions essentially depends on the system parameters 
characterizing different prevailing situations, when the main features are 
described as gain and competition (Sec. 4), loss and cooperation (Sec. 5), loss
and competition or gain and cooperation (Sec. 6). The equations display a rich 
variety of dynamical regimes, including punctuated unbounded growth, punctuated 
increase or punctuated degradation to a stationary state, convergence to a 
stationary state with sharp reversals of plateaus, oscillatory convergence, 
everlasting fluctuations, everlasting up-down plateau reversals, and divergence 
in finite time. All admissible dynamical regimes are studied and illustrated. The 
role of noise on the dynamics is investigated in Sec. 7. The possibility of 
predicting finite-time singularities by observing only the initial stage of 
motion is discussed in Sec. 8. Section 9 concludes.

\section{Evolution equation with functional delayed carrying capacity}

Here and in what follows, we use the dimensionless variable for the population
$x(t)$ as a function of time $t$. The reduction of the dimensional equation to
the dimensionless form has been explained in full details in our previous papers
\cite{Yukalov_13,Yukalov_14} and we do not repeat it here.

\subsection{Singular solutions for linear delayed carrying capacity}

The general expression for the evolution equation with functional carrying 
capacity, in dimensionless notation, reads as
\be
\label{1}
 \frac{dx(t)}{dt} = \sgm_1 x(t) -\sgm_2 \; \frac{x^2(t)}{y(x)} \;  ,
\ee
where the carrying capacity functional
\be
\label{2}
y(x) = y[x(t-\tau)]
\ee
depends on the population at an earlier time, with a constant delay time $\tau$,
which embodies that any influence of the population on the capacity requires 
time in order to either create additional means or to destroy the given resources. 
Here and in what follows, we use the term {\it population}, although the variable 
$x$ can characterize either population, or firm assets, or other financial and 
economic indices \cite{Yukalov_13,Yukalov_14}.

The coefficients $\sigma_i$ describe the prevailing features in the balance between
gain (birth) or loss (death) and competition versus cooperation. There exist four
situations characterized by these coefficients:
\begin{eqnarray}
\label{3}
\begin{array}{lll}
\sgm_1 = 1 \; ,  & ~ \sgm_2 = 1  & ~~~~~ (gain \; \& \; competition) \; , \\
\sgm_1 = -1 \; , & ~ \sgm_2 = -1 & ~~~~~ (loss \; \& \; cooperation) \; , \\
\sgm_1 = -1 \; , & ~ \sgm_2 = 1  & ~~~~~ (loss \; \& \; competition) \; , \\
\sgm_1 = 1 \; ,  & ~ \sgm_2 = -1 & ~~~~~ (gain \; \& \; cooperation) \; .
\end{array}
\end{eqnarray}
Various admissible interpretations of the equation and possible applications 
have been described in the published papers \cite{Yukalov_13,Yukalov_14}.

If we fix $\sigma_1 = \sigma_2 = 1$ and take a constant capacity $y_0 = y(0)$, 
we come back to the logistic equation. In our previous papers 
\cite{Yukalov_13,Yukalov_14}, we modelled the influence of population on the 
carrying capacity by the linear approximation
\be
\label{4}
 y_1(x) = 1 + b_1 x(t-\tau) \;   ,
\ee
with the parameter $b_1$ describing either destructing action of population on 
the resources, when $b_1 < 0$, or creative population activity, if $b_1 > 0$.

In the case of destructive action, it may happen that the capacity (4) reaches 
zero and becomes even negative, with the effect of the growing population. It 
crosses zero at a critical time $t_c$ defined by the equation
$$
 1 + b_1 x(t_c-\tau) = 0 \;  .
$$
Being in the denominator of the second term of (1), the vanishing capacity leads 
to the appearance of divergent or non-smooth solutions. In some cases, having to 
do with financial and economic applications, the arising negative capacity can 
be associated with the leverage effect \cite{Yukalov_14}. However, in the usual 
situation, the solution divergencies, caused by the zero denominator, look 
rather unrealistic, reminding of mathematical artifacts. Therefore, it would 
be desirable to define a carrying capacity that would not cross zero in finite 
time.

\subsection{Evolution equation with nonlinear delayed carrying capacity}

It would be possible to replace the linear form (\ref{4}) by some nonlinear 
function. This, however, would be a too arbitrary and ambiguous procedure, since
it would be always unclear why this or that particular function has been chosen.
In order to justify the choice of a nonlinear function, we propose
the following procedure to select the form of the nonlinear carrying capacity.   

Strictly speaking, the linear approximation presupposes that the second term 
in the r.h.s of expression (4) is smaller than one. Generally, a function 
$y(x)$ can be expanded in powers of its variable $x$ according to the series
\be
\label{5}
 y(x) \simeq 1  + b_1 x + b_2 x^2 + \ldots \;  ,
\ee
where different terms describe the influence with different action intensity.
Since expression (\ref{4}) can be interpreted as the first-order term in the 
general series expansion (\ref{5}) with, in principle, infinite many terms, 
it is convenient to think of it as the general expansion of some nonlinear 
function to be determined by a suitable summation. We thus propose to construct 
a nonlinear extension of (\ref{4}) by defining an effective sum of these 
series (\ref{5}). A standard way to realize the summation (\ref{5}) is via 
Pad\'{e} approximants \cite{Baker_15}. However, as is well known, the Pad\'{e}
approximants are very often plagued by the occurrence of artificial zeroes and
divergencies, which makes them inappropriate for the summation of a quantity that 
is required to be finite and positive. For this purpose, it is more appropriate 
to resort to the method of self-similar approximations
\cite{Yukalov_16,Yukalov_17,Yukalov_18,Yukalov_19,Yukalov_20}. This is a 
mathematical method allowing for the construction of effective sums of power-law 
series, even including divergent series. According to this method, a series 
(\ref{5}) is treated as a trajectory of a dynamical system, whose fixed point 
represents the sought effective sum of the series. In the vicinity of a fixed
point, the trajectory becomes self-similar, which gives its name to the method
of self-similar approximations.  

Employing this method, under the restriction of getting a positive effective 
sum of the series, which is done by using the self-similar exponential 
approximants \cite{Yukalov_21,Yukalov_22}, we obtain the effective sum
\be
\label{6}
 y(x) = \exp\{ b x(t-\tau) \} \;  .
\ee
Here the {\it production parameter} $b$ characterizes the type of the influence
of the population on the carrying capacity. In the case of creative activity of
population, producing additional means for survival, the creation parameter is
positive, $b > 0$. And if the population destroys the given carrying capacity,
then the destruction parameter is negative ($b < 0$).

In this way, we come to the evolution equation
\be
\label{7}
 \frac{dx(t)}{dt} = \sgm_1 x(t) - \sgm_2 x^2(t) \exp \{ - bx(t-\tau) \} \; ,
\ee
with the nonlinear carrying capacity that generalizes the delayed logistic 
equation (\ref{1}) with the linear capacity. The proposed generalization (\ref{7}) 
allows us to consider population influences of arbitrarily strong intensity. 
The initial condition to the equation defines the history
\be
\label{8}
 x(t) = x_0 \qquad ( t \leq 0 ) \;  .
\ee
By definition, the population is described by a positive variable, so that we 
will be looking for only positive solutions $x(t) > 0$ for $t > 0$.

\section{Existence and stability of stationary states}

This section gives the general conditions for the existence and stability of 
stationary states of the delayed Eq. (7). The details of such conditions depend 
on the type of the system characterized by the values of $\sigma_i$. More 
specific investigations of the evolutionary stable states as well as the overall 
dynamical regimes will be analyzed in the following sections.

\subsection{Existence of stationary states}

The stationary states of Eq. (7) are defined by the solutions to the equation
\be
\label{9}
 \sgm_1 x^* - \sgm_2 \left ( x^* \right )^2 \exp ( - bx^*) = 0 \; .
\ee
There always exists the trivial solution
\be
\label{10}
 x_1^* = 0 \qquad
(-\infty < b < \infty\; , ~ \sgm_1=\pm 1 \; , ~ \sgm_2 =\pm 1 )\; .
\ee
But the nontrivial solutions require the validity of the relation
\be
\label{11}
 \frac{\sgm_1}{\sgm_2} = x^* \exp(-bx^* ) \;  ,
\ee
which tells us that they may happen only for
\be
\label{12}
\sgm_1 = \sgm_2 = \pm 1  \qquad ( x^* > 0 ) \; .
\ee
Two nontrivial states may exist depending on the value of the production 
parameter. One of the states exists for negative values of the latter in the 
range
\be
\label{13}
 0 < x_2^* \leq 1 \qquad (b\leq 0 ) \;  .
\ee

Below we show that varying the parameters of Eq. (7) generates a number of 
bifurcations and provides a rich variety of qualitatively different solutions. 
We give a complete classification of all possible solutions demonstrating how 
the bifurcation control \cite{Chen} can be realized for this equation.

For positive $b$, there can occur two states, such that
\be
\label{14}
 1 < x_2^* < e \; , \qquad x_3^* > e \qquad
\left ( 0 < b < \frac{1}{e} \right ) \;  .
\ee
At the bifurcation point $b = 1/e$, the states coincide: $x_2^* = x_3^* = e$.
There are no stationary nontrivial states for $b > 1/e$. The existence of the
nontrivial states is illustrated in Fig. 1.

\begin{figure}[ht]
\centerline{\includegraphics[width=6.5cm]{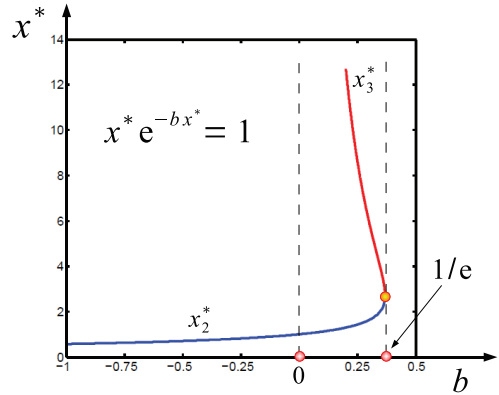}}
\caption{Existence of nontrivial stationary states depending on the value 
of the production parameter $b$.} 
\label{fig:Fig.1}
\end{figure}

\subsection{Stability of stationary states}

Studying the stability of the stationary states for the delay differential
equation, we follow the Lyapunov-type procedure developed for delay equations
in the book \cite{Kolmanovskii_23} and described in detail in Refs.
\cite{Yukalov_13,Yukalov_14}. The steps of this procedure are as follows. We
consider a small deviation from the solution of equation (7) by writing 
$x = x^* + \delta x$. Substituting this in (7) yields the linearized equation
\be
\label{15}
\frac{d}{dt} \; \dlt x(t) = \left ( \sgm_1 - 2\sgm_2 x^* e^{-bx^*} \right )
\dlt x(t) + b \sgm_1 x^*\dlt x(t-\tau)  \;  ,
\ee
which is the basis for the stability analysis. The solution of equation (7)
is Lyapunov stable, when the solution to (15) is bounded. A fixed point is 
asymptotically stable, when the solution to (15) converges to zero, as time
tends to infinity. Thus, the conditions of fixed-point stability are prescribed
by the convergence to zero of the small deviation described by equation (15).      

We find that the trivial state $x_1^* = 0$ is stable when
\be
\label{16}
 \sgm_1 = -1 \; , \qquad \sgm_2 = \pm 1 \; , \qquad
- \infty < b < \infty \; , \qquad \tau \geq 0 \;  ,
\ee
while for $\sigma_1 = 1$ it is always unstable.

In the case of the nontrivial states, we use relation (11) and reduce Eq. (15) 
to
\be
\label{17}
 \frac{d}{dt} \; \dlt x(t) = - \sgm_1 \dlt x(t) + 
b \sgm_1 x^* \dlt x(t-\tau) \; .
\ee
According to the existence condition (12), we need to study the stability 
of the nontrivial states only for coinciding $\sigma_i$.

The following analysis of the stationary states and the solution of the full
evolution equation (7) requires to specify the values of $\sigma_i$.

\section{Dynamics under gain and competition ($\sgm_1 = \sgm_2 = 1$)}

\subsection{Evolutionary stable states}

The stability analysis shows that the nontrivial state $x_2^*$ is stable either 
in the range
\be
\label{18}
0 < x_2^* \leq \frac{1}{e} \qquad  ( b \leq -e \; , ~~ \tau <\tau_2^* )
\ee
or in the range
\be
\label{19}
 \frac{1}{e} < x_2^* \leq e \qquad \left ( - e < b \leq \frac{1}{e} \; , ~~
\tau > 0 \right ) \; ,
\ee
where
\be
\label{20}
 \tau_2^* = \frac{1}{\sqrt{(bx_2^*)^2-1} } \;
\arccos\left ( \frac{1}{bx_2^*} \right ) \;  .
\ee
Close to the boundary $b \ra - e$ one has
$$
x_2^* \simeq \frac{1}{e} \left ( 1 + \frac{b+e}{2e} \right ) \; , \qquad
\tau_2^* \simeq \frac{\pi\sqrt{e}}{|b|-e} \qquad (b\ra -e) \;   .
$$
The state $x_3^* > e$ is always unstable under $\sigma_1 = \sigma_2 = 1$.

Thus, there can exist just one stable stationary state $x_2^* \in (0,e)$, 
whose region of stability is given by Eqs. (\ref{18}) to (\ref{20}) and shown 
in Fig. 2.

\begin{figure}[ht]
\centerline{\includegraphics[width=6.5cm]{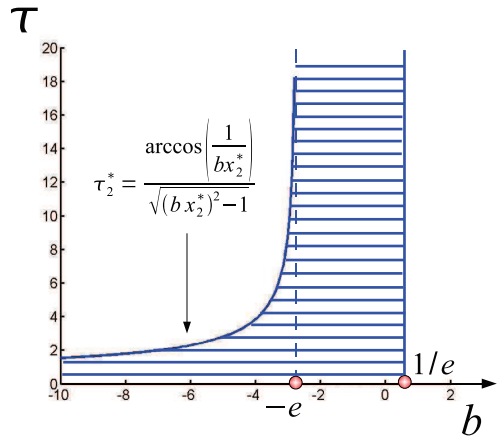}}
\caption{Stability region (shadowed) for the stationary state $x_2^*$ under
$\sigma_1 = \sigma_2 = 1$.} 
\label{fig:Fig.2}
\end{figure}

For some values of the production parameter $b$, the basin of attraction of
$x_2^*$ is not the whole positive semiline of $x_0$, but a limited interval.
This happens for $b \in (0, 1/e)$, when the basin of attraction is defined by 
the inequalities
\be
\label{21}
 0 < x_0 < x_3^* \qquad \left ( 0 < b \leq \frac{1}{e}  \right ) \; .
\ee
For these $b$ values, the solution $x$ tends to infinity, as $t \ra \infty$, 
for the history $x_0 > x_3^*$, even for the parameters $b$ and $\tau$ in the 
region of stability of $x_2^*$.

\subsection{Boundedness of solutions for semi-negative production parameters}

When $\sgm_1 = \sgm_2 = 1$ and when the production parameter $b$ is non-positive,
this means that the population does not produce its carrying capacity but rather 
destroys it or, in the best case, retains the given capacity value. In this case, 
an important result for the overall temporal behavior of solutions can be derived
rigorously: the {\it population growth has to be limited}.

\vskip 2mm
\begin{proposition}
The solution $x(t)$ to the evolution equation (7), under the condition 
$\sgm_1 = \sgm_2 = 1$, for $b \leq 0$, any finite $\tau \geq 0$, and any history
$x_0 \geq 0$, is bounded for all times $t \geq 0$, and, for $b < 0$, there exists 
a time
$t_0 = t_0(x_0,\tau)$ such that
\be
\label{22}
0 \leq x(t) \leq 1 \qquad (t\geq t_0 )\;  .
\ee
\end{proposition}
\vskip 2mm

\begin{proof}
When $b = 0$, the explicit solution is
$$
 x(t) = \frac{x_0}{x_0+(1-x_0)e^{-t} } \qquad ( b = 0 ) \;  .
$$
If $x_0 < 1$, then $x \ra 1$ from below as $t \ra \infty$. If $x_0 = 1$, 
then $x = 1$ for all $t > 0$. If $x_0 > 1$, then $x \ra 1$ from above, as 
$t\ra\infty$. So, the solution is always bounded.

When $b$ is arbitrary and $t \leq \tau$, then the explicit solution
$$
x(t) = \frac{x_0e^{bx_0+t}}{e^{bx_0}+x_0(e^t-1)}
$$
is evidently bounded.

For $b < 0$, any finite $\tau \geq 0$ and all $t \geq 0$, the evolution 
equation (7) reads as
\be
\label{23}
\frac{dx}{dt} = x - x^2 \exp \{ | b | x(t - \tau) \} \;  .
\ee
If, at some moment of time $t > 0$, it happens that $x \geq 1$, then the above 
equation defines a semi-negative derivative
$$
 \frac{dx}{dt} \leq x ( 1 - \exp\{ | b | x(t - \tau) \} ) \leq 0 \; ,
$$
implying that $x$ decreases or does not grow.

If the history is such that $x_0 < 1$, then either $x$ stays always below one, 
or it grows and reaches one at some moment of time $t > 0$. But $x$ cannot cross 
the line $x = 1$, since, as is shown above, at the time when $x$ would become 
$\geq 1$, it has to either stay on this line $x = 1$ or has to decrease. The 
solution cannot stay forever on the unity line, as far as $x = 1$ is not a stable
stationary state, which is $x_2^* < 1$ for $b < 0$. This means that there is a 
moment of time $t_0 < +\infty$ such that the solution $x$ has to go down for 
$t>t_0$.

For $x_0 = 1$, again the solution cannot rise, having a non-positive derivative, 
and cannot stay forever at this value, which is not a stable fixed point.
The sole possibility is that $x$ starts diminishing beyond a finite $t_0$.

When $x_0 > 1$, then its derivative is non-positive. The solution cannot grow and 
cannot stay forever at the value that is larger than the fixed point $x_2^* < 1$. 
Hence $x$ must decrease. When diminishing, it reaches the value $x = 1$, where 
it cannot stay forever, but has to go down beyond a finite time $t_0$.

Thus, for $b < 0$, there always exists such a moment of time, when the solution 
goes below the value $x = 1$ and can never cross this line from below. 
\end{proof}

\subsection{Punctuated unbounded growth}

A different situation happens for positive production parameters, when the
solutions may be unbounded. For example, when $b$ is outside of the stability
region of $x_2^*$, so that
\be
\label{24}
b > \frac{1}{e} \; , \qquad \tau \geq 0 \; , \qquad x_0 > 0 \;   ,
\ee
then $x$ tends by steps to infinity, as $t \ra \infty$. A similar unbounded
punctuated growth occurs when $b$ is inside the stability region, but $x_0$ 
is outside of the attraction basin of $x_2^*$, which takes place if
\be
\label{25}
 0 < b < \frac{1}{e} \; , \qquad \tau \geq 0 \; , \qquad x_0 > x_3^* \;  .
\ee
This behavior is illustrated by Fig. 3.

\begin{figure}[ht]
\centerline{\includegraphics[width=6.5cm]{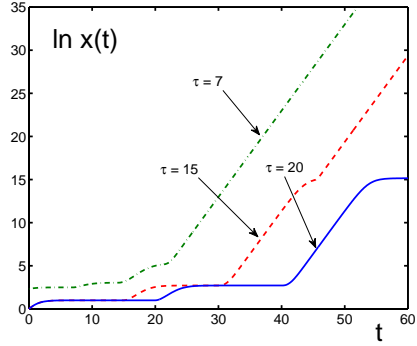}}
\caption{Punctuated unbounded growth, under $\sigma_1=\sigma_2=1$, for the 
parameters: $b = 1 > 1/e$, $\tau = 20$, $x_0 = 1$ (solid line); 
$b = 1$, $\tau = 15$, $x_0 = 1$ (dashed line); 
$b = 0.25$, $\tau = 7$, $x_0 = 10 > x_3^* = 8.613$ (dashed-dotted line).} 
\label{fig:Fig.3}
\end{figure}

The meaning of such a behavior is clear: for either sufficiently high production
parameter or for sufficiently high startup and creative activity, the population
(or firm development) can demonstrate unlimited punctuated growth with time.

\subsection{Punctuated convergence to stationary state}

For positive production parameters, the solutions can also tend to the 
stationary state $x_2^*$ by punctuated steps. They tend to $x_2^*$ from below 
if $x_0 < x_2^*$, and from above if $x_0 > x_2^*$. This happens when
\be
\label{26}
0 < b < \frac{1}{e} \; , \qquad \tau \geq 0 \; , \qquad x_0 < x_3^* \;   .
\ee

When the production parameter is negative, the approach to the stationary 
state becomes quite nonmonotonic, with sharp reversals after almost horizontal 
plateaus. This regime arises when
\be
\label{27}
 - e < b < 0 \; , \qquad \tau \geq 0 \; , \qquad x_0 > 0 \;   .
\ee
With decreasing further the negative destruction parameter, the plateaus shorten
and the dynamics reduce to strongly fluctuating convergence to the focus $x_2^*$,
which occurs for
\be
\label{28}
  b < -e  \; , \qquad \tau < \tau_2^* \; , \qquad x_0 > 0 \;  ,
\ee
where the critical time lag is given by (\ref{20}).

Different regimes of punctuated convergence to the stationary state $x_2^*$ 
are shown in Fig. 4, which include punctuated growth, punctuated decay, and 
convergence with plateau reversals. Strongly oscillatory convergence to $x_2^*$ 
is demonstrated in Fig. 5.

\begin{figure}[ht]
\vspace{9pt}
\centerline{
\hbox{ \includegraphics[width=6.5cm]{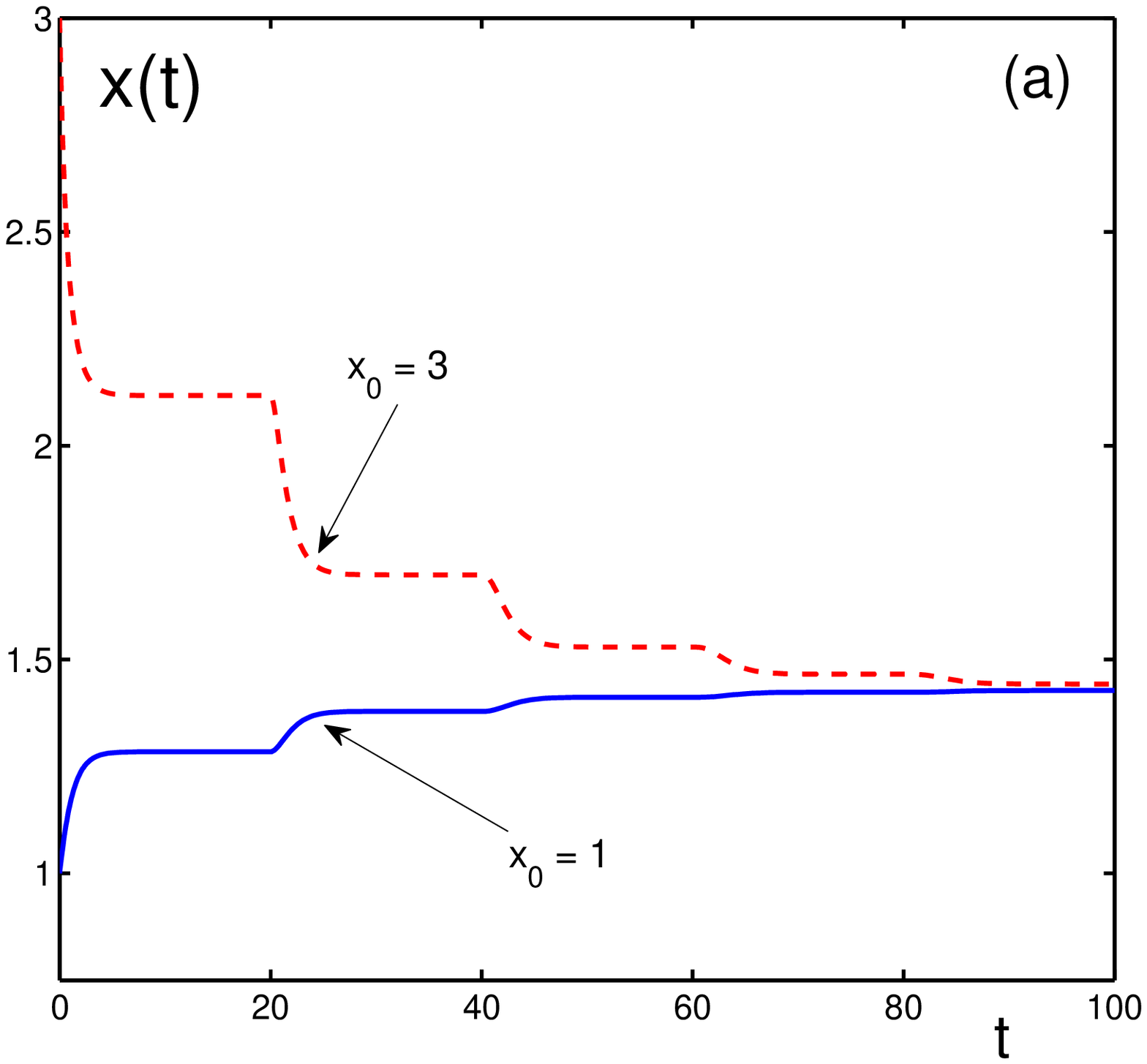} \hspace{2cm}
\includegraphics[width=6.5cm]{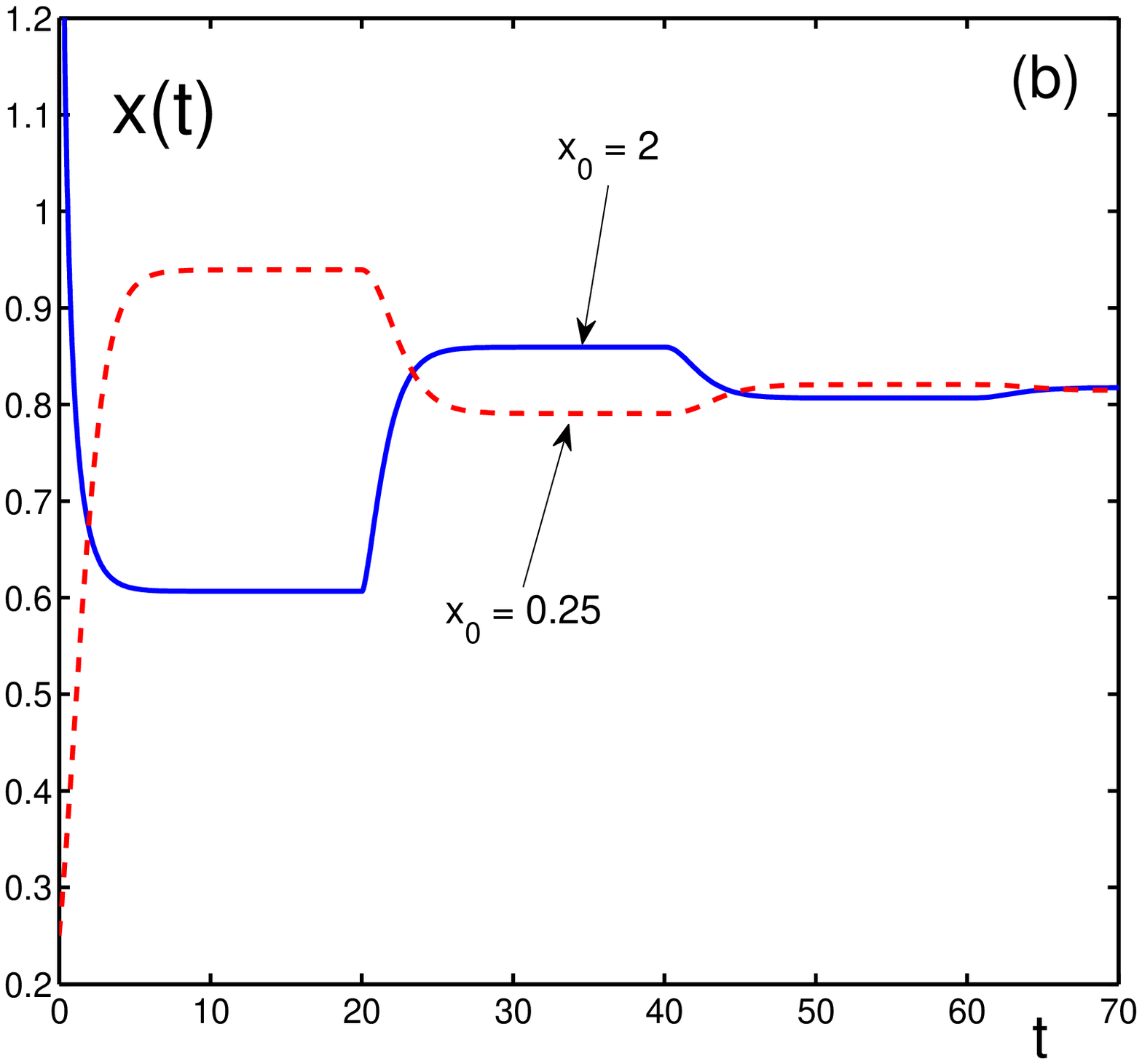} } }
\caption{Different types of punctuated convergence to the stationary state $x_2^*$, 
under $\sigma_1 = \sigma_2 = 1$, for different parameters: (a) punctuated growth
for $b = 0.25$, $\tau = 20$, $x_0 = 1 < x_2^* = 1.43$ (solid line); punctuated decay 
for $b = 0.25$, $\tau = 20$, $x_0 = 3 > x_2^* = 1.43$ (dashed line); (b) convergence 
with plateau reversals under negative $b = -0.25 > -e$ and $\tau = 20$ for 
$x_0 = 2 > x_2^* = 0.816$ (solid line) and $x_0 = 0.25 < x_2^* = 0.816$ (dashed line).}
\label{fig:Fig.4}
\end{figure}

\begin{figure}[ht]
\vspace{9pt}
\centerline{
\hbox{ \includegraphics[width=6.5cm]{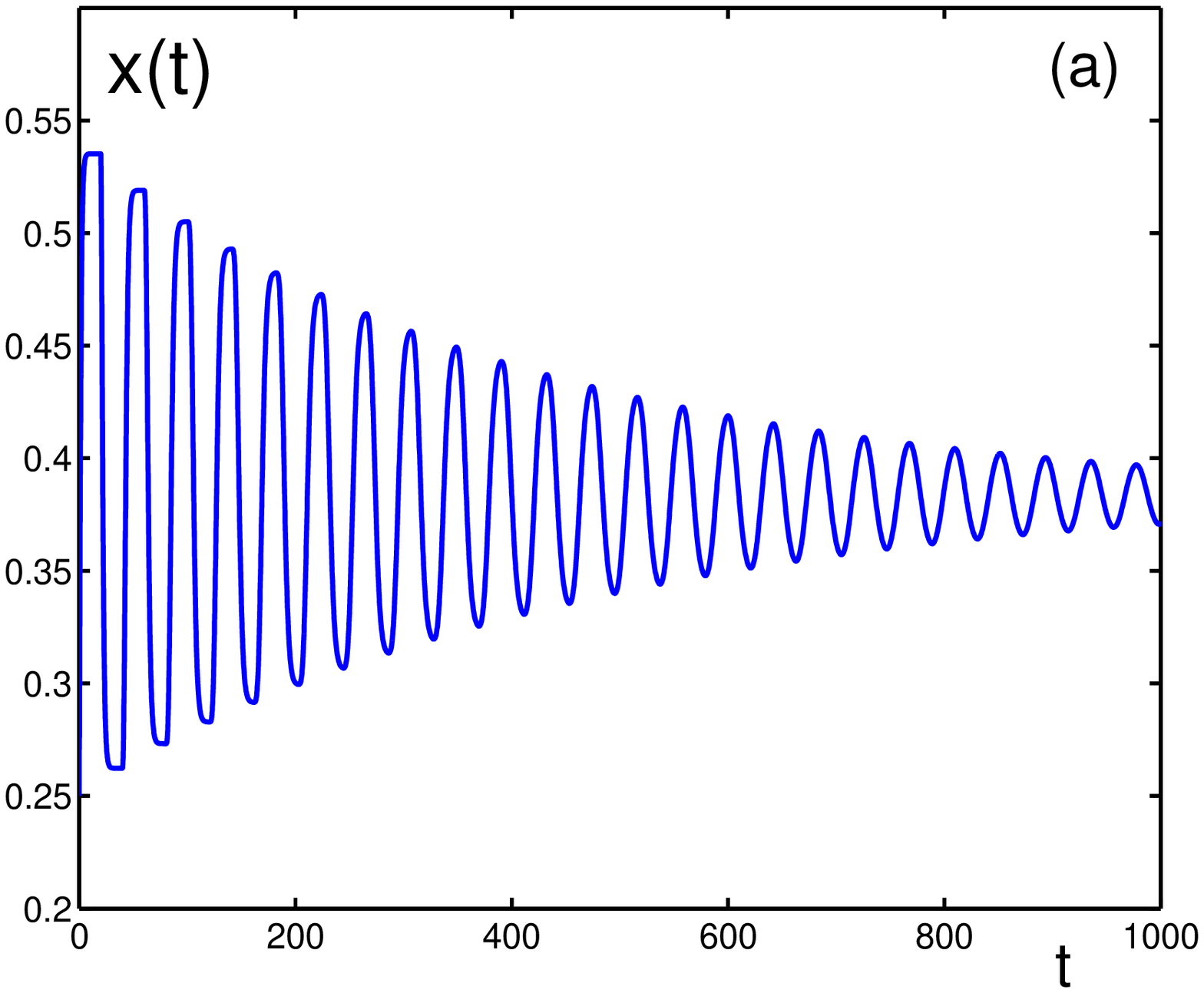} \hspace{2cm}
\includegraphics[width=6.5cm]{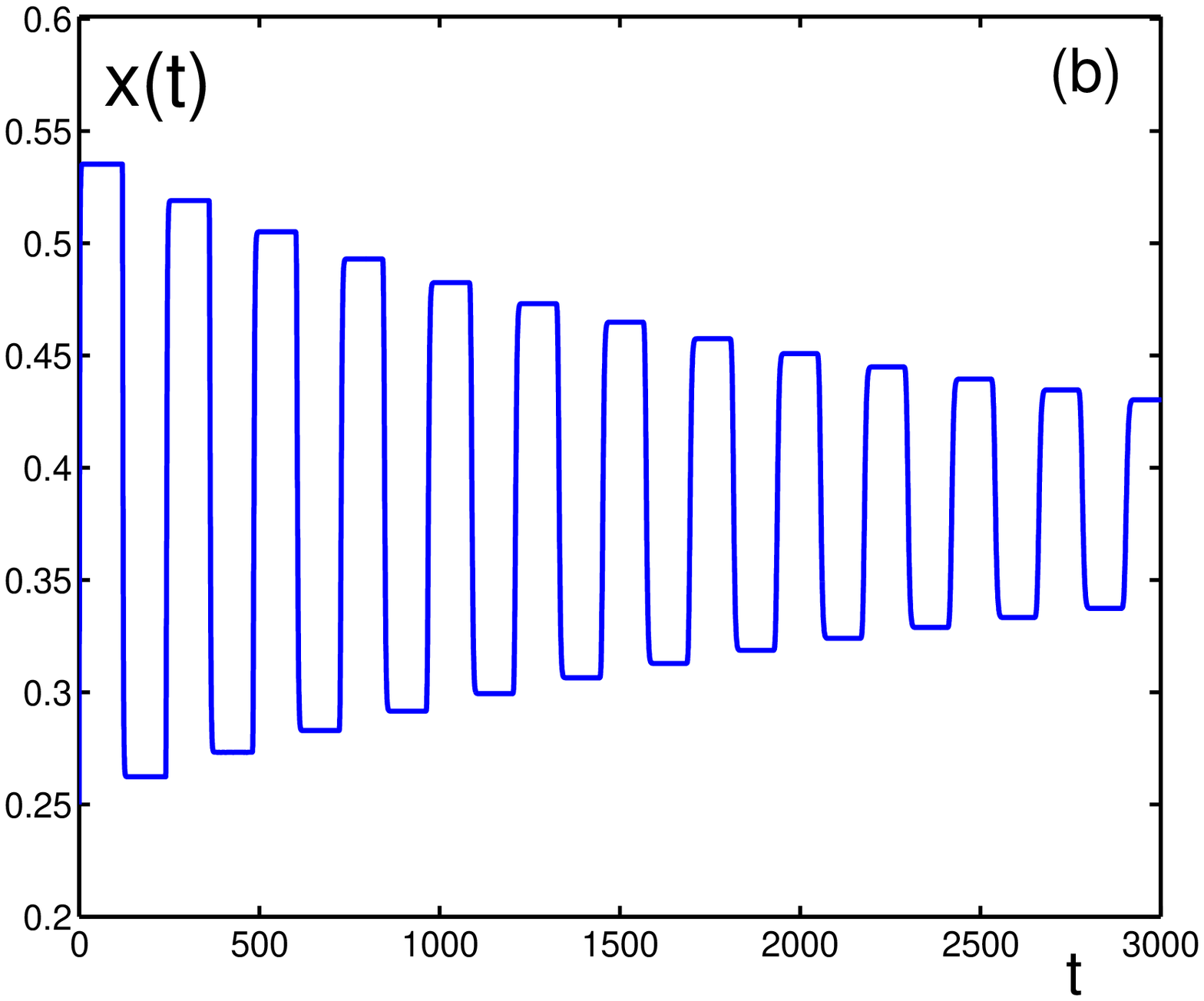} } }
\caption{Strongly oscillatory convergence to the stationary state $x_2^* = 0.383$, 
under $\sigma_1 = \sigma_2 = 1$, with $b = -2.5 < -e$ and $x_0 = 0.25$, for different
time lags: (a) $\tau = 20$; (b) $\tau = 120$.}
\label{fig:Fig.5}
\end{figure}

\subsection{Everlasting oscillations}

When the destructive action is rather strong, increasing the time delay leads 
to the switch from the oscillatory convergence to a stationary state to the 
regime of everlasting oscillations, which happens for
\be
\label{29}
  b < -e  \; , \qquad \tau \geq \tau_2^* \; , \qquad x_0 > 0 \;  .
\ee
Figure 6 illustrates this effect of changing the dynamical regime with increasing 
the time lag above $\tau_2^*$ that plays the role of a critical point.

\begin{figure}[ht]
\vspace{9pt}
\centerline{
\hbox{ \includegraphics[width=6.5cm]{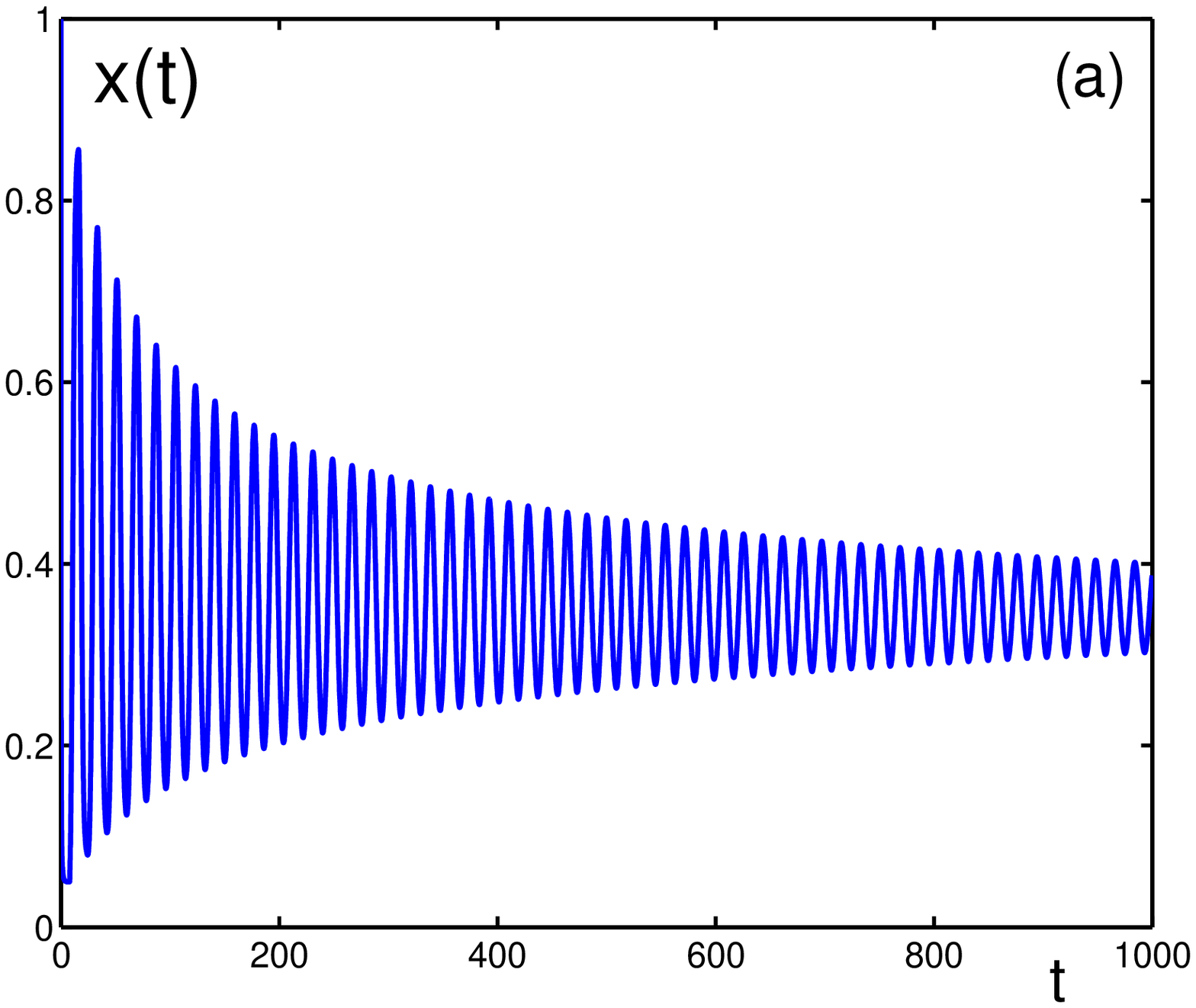} \hspace{2cm}
\includegraphics[width=6.5cm]{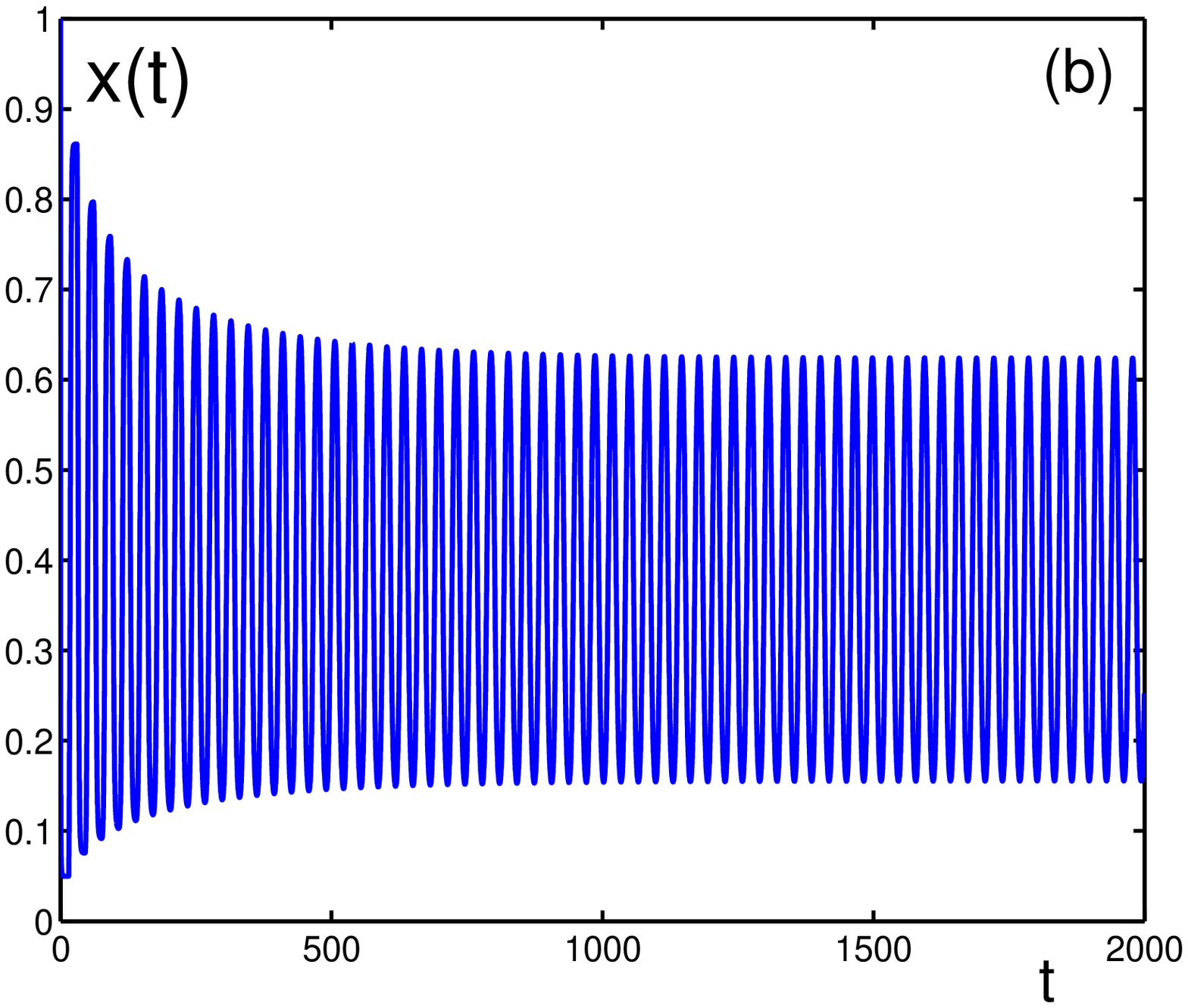} } }
\vspace{9pt}
\centerline{
\hbox{ \includegraphics[width=6.5cm]{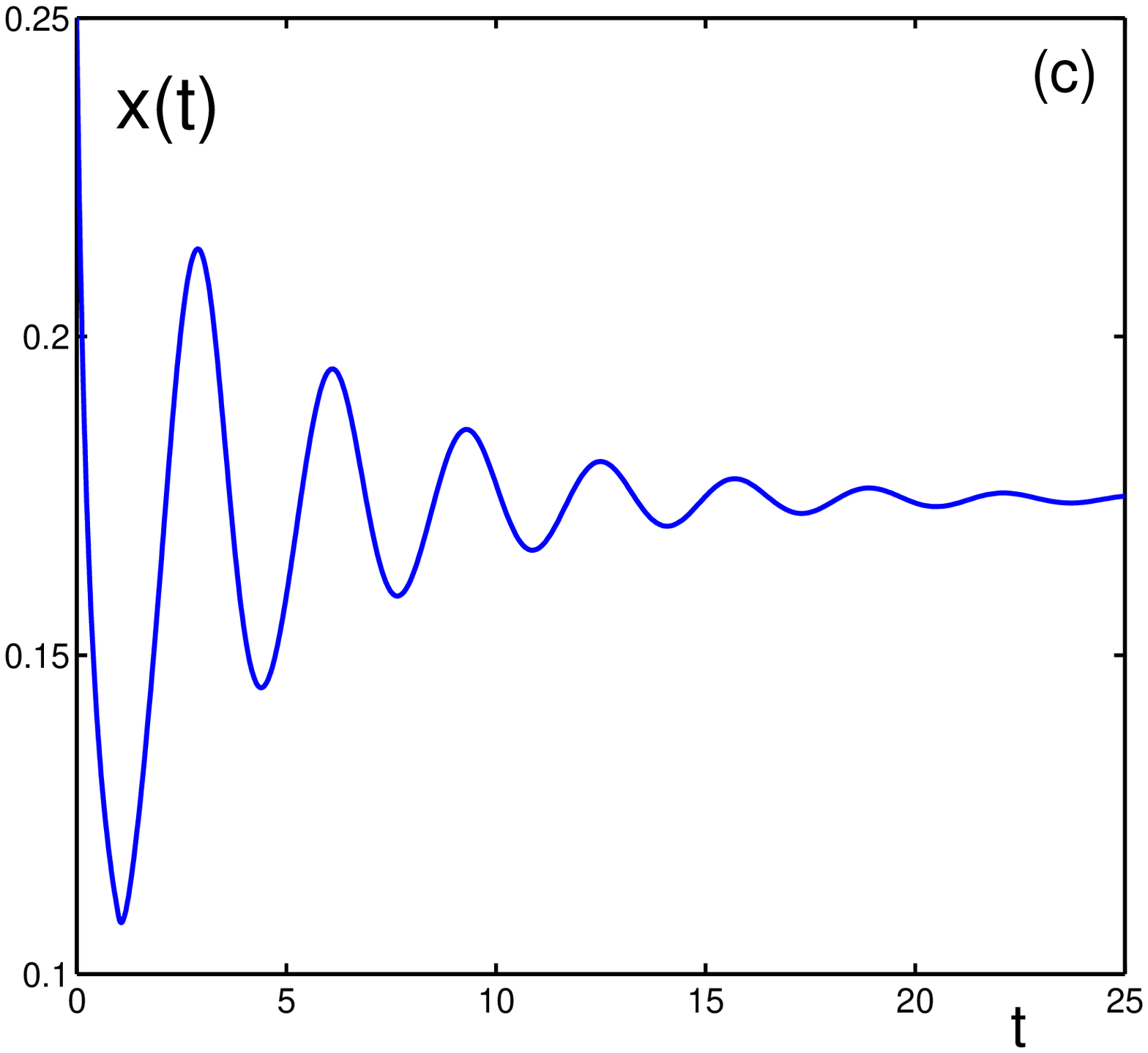} \hspace{2cm}
\includegraphics[width=6.5cm]{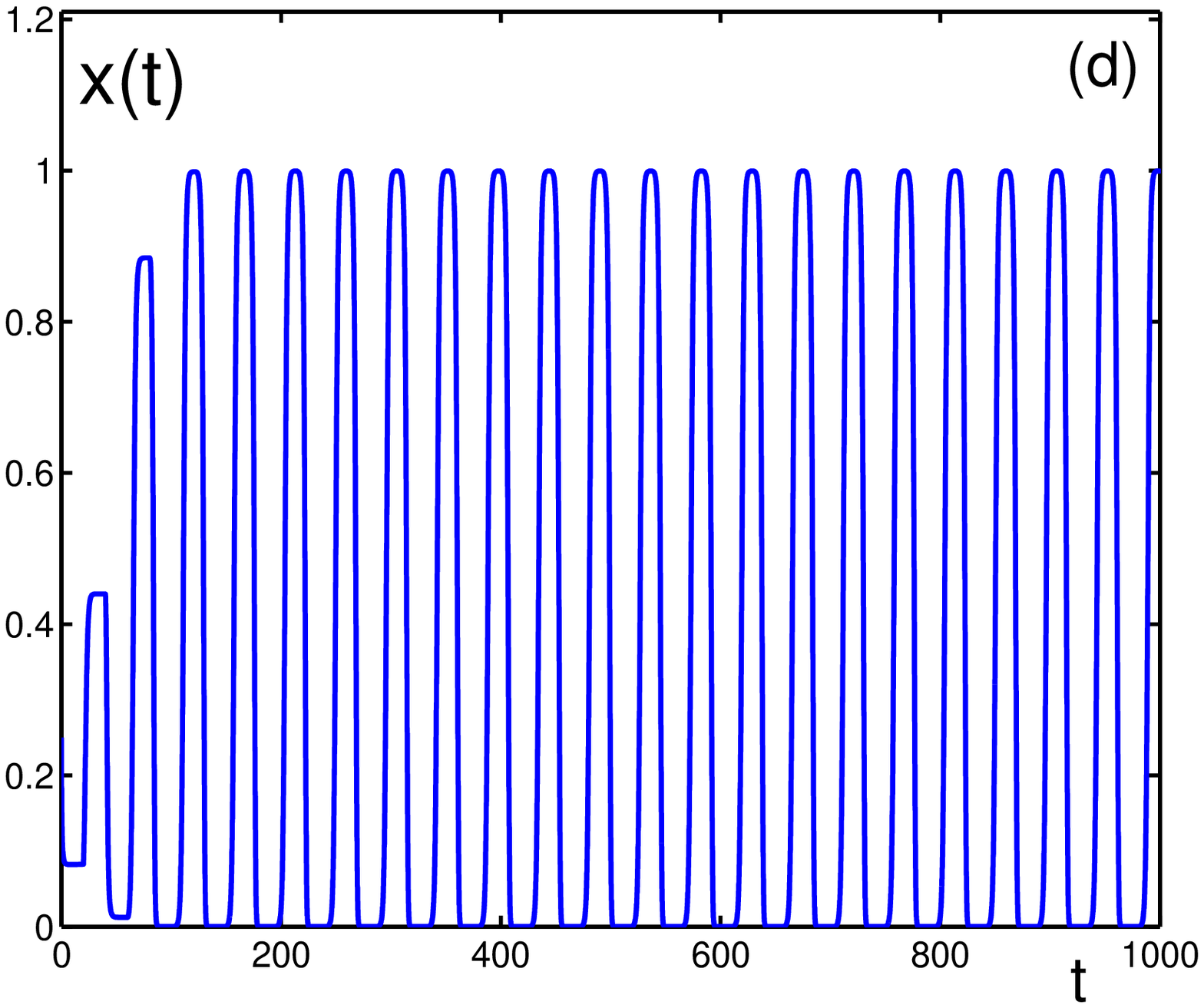} } }
\caption{Regime switch from fluctuating convergence to everlasting oscillations, 
under $\sigma_1 = \sigma_2 = 1$, with increasing the time lag over its critical 
value:
(a) fluctuating convergence to $x_2^* = 0.35$ for 
$b = -3 < -e$, $x_0 = 1$, $\tau = 8 < \tau_2^* = 8.854$; 
(b) everlasting oscillations 
for $b = -3$, $x_0 = 1$, $\tau=15 > \tau_2^* = 8.854$; 
(c) fluctuating convergence to
$x_2^*=0.175$ for $b =-10 <-e$, $x_0=0.25$, $\tau=1<\tau_2^*=1.524$; 
(d) 
everlasting oscillations for $b =-10$, $x_0=0.25$, $\tau=20>\tau_2^*=1.524$.}
\label{fig:Fig.6}
\end{figure}

A rather exotic regime develops when the feedback action on the carrying 
capacity is strongly destructive and the time lag is very long. Then, there 
appears the regime of everlasting up-down reversals of plateaus located at zero 
and one, as is shown in Fig. 7.

\begin{figure}[ht]
\centerline{\includegraphics[width=6.5cm]{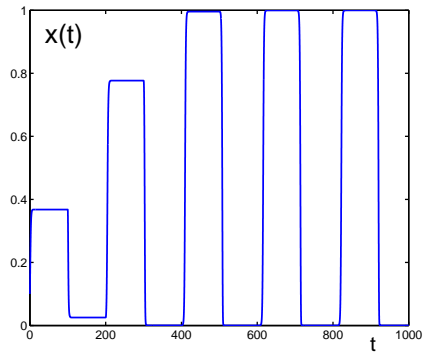}}
\caption{Everlasting up-down plateau reversals, $\sigma_1=\sigma_2=1$, 
for the destruction parameter $b = -10 \ll -e$, history $x_0=0.1$, and 
the time lag $\tau = 100 \gg \tau_2^* = 1.524$.} 
\label{fig:Fig.7}
\end{figure}

\section{Dynamics under loss and cooperation ($\sgm_1 = \sgm_2 = -1$)}

\subsection{Existence of finite-time singularities}

Contrary to the previous case of gain and competition, now there can appear 
unbounded solutions diverging at a finite time. For instance, in the region 
$t < \tau$, there exists the critical time
\be
\label{30}
 t_c = - \ln \left ( 1 -\; \frac{e^{bx_0}}{x_0} \right ) \;  ,
\ee
where $x$ hyperbolically diverges, as $t \ra t_c$, for some $b$ and $x_0$. 
The relation between $b$ and $x_0$, for which the divergence occurs, is shown 
in Fig. 8.

\begin{figure}[ht]
\centerline{\includegraphics[width=8cm]{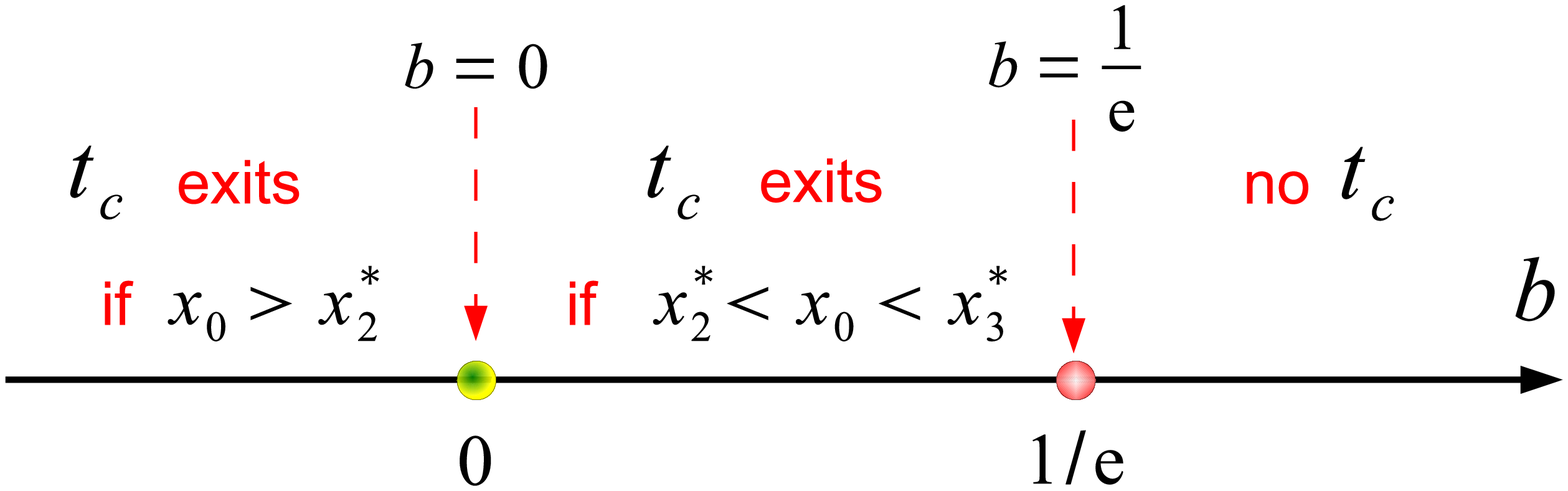}}
\caption{Conditions for the existence of singular solutions under 
$\sigma_1 = \sigma_2 = -1$, depending on the relation between $b$ 
and $x_0$.} 
\label{fig:Fig.8}
\end{figure}

The occurrence of a finite-time singularity is associated with the development
of an instability of the system, and the critical time $t_c$ corresponds to 
the time of a change of regime. Concrete interpretations for various dynamical 
systems with such singularities, corresponding to population growth, mechanical 
ruptures or fractures, and economic or financial bubbles, have been discussed 
in detail in many articles 
\cite{Kapitza_1,Hern_2,Korotayev_3,Yukalov_28,Yukalov_13,Yukalov_14,Johansen_24,
Fogedby_25,Sornette_26,Sornette_30,Andersen_27,Andersen_29}.

\subsection{Evolutionary stable states}

Under loss and cooperation, the state $x_2^*$ is always unstable. But
the trivial stationary state $x_1^* = 0$ is also stable for all $b$ and $\tau$. 
The nontrivial state $x_3^* > e$ is stable for
\be
\label{31}
 0 < b < \frac{1}{e} \; , \qquad \tau < \tau_3^* \;  ,
\ee
where
\be
\label{32}
  \tau_3^* = \frac{1}{\sqrt{(bx_3^*)^2-1} } \;
\arccos \left ( \frac{1}{bx_3^*} \right ) \; .
\ee
On the boundary of stability, when $b$ approaches $1/e$, then
$$
 x_3^* \simeq e [ 1 + \sqrt{2(1-be)} ] \; , \qquad
\tau_3^* \simeq 1 - \; \frac{2}{3} \; \sqrt{2(1-be)} \qquad \left (
b \ra \frac{1}{e} \right ) \; .
$$
The stability region of $x_3^*$ is presented in Fig. 9. Because the state 
$x_1^*$ is stable everywhere, hence the stability region of $x_3^*$ corresponds 
to the region of bistability.

\begin{figure}[ht]
\centerline{\includegraphics[width=6.5cm]{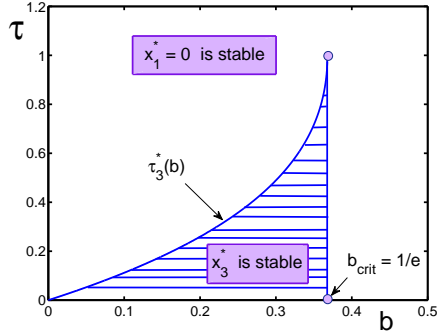}}
\caption{Stability region of the stationary state $x_3^*$ (shadowed), 
under $\sigma_1 = \sigma_2 = -1$. Since $x_1 = 0$ is always stable, 
the shadowed region is also the region of bistability.} 
\label{fig:Fig.9}
\end{figure}

\subsection{Dynamical regimes of evolution}

In the case of loss and cooperation, depending on the parameters $b$, $\tau$, 
and the history $x_0$, there can occur the following dynamical regimes: monotonic 
convergence to a stationary state, convergence with oscillations, everlasting 
oscillations, and finite-time singularity. All these regimes are demonstrated 
in Figs. 10 to 13. The summary of all possible solution types is illustrated 
in the scheme of Fig. 14.

\begin{figure}[ht]
\vspace{9pt}
\centerline{
\hbox{ \includegraphics[width=6.5cm]{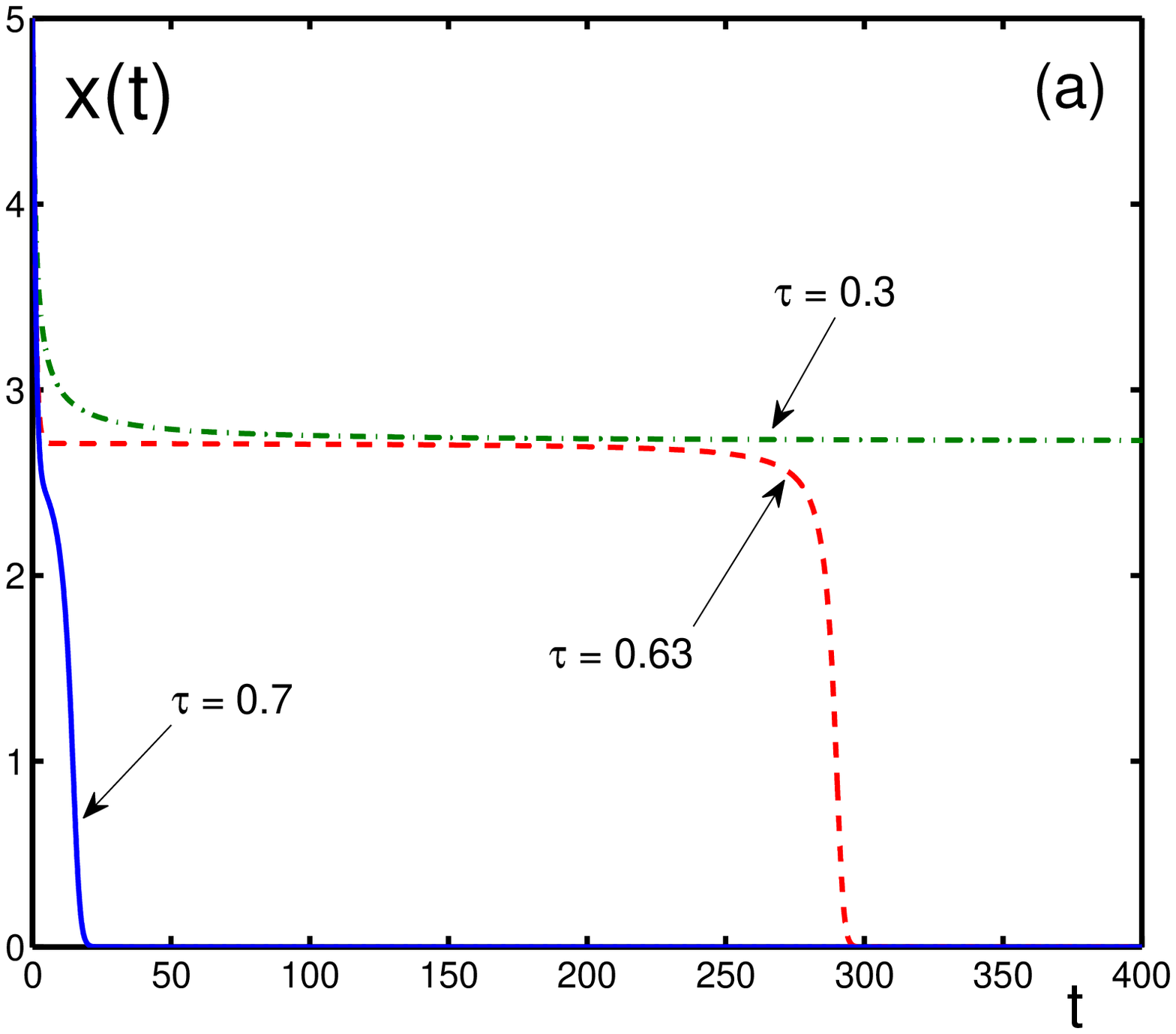} \hspace{2cm}
\includegraphics[width=6.5cm]{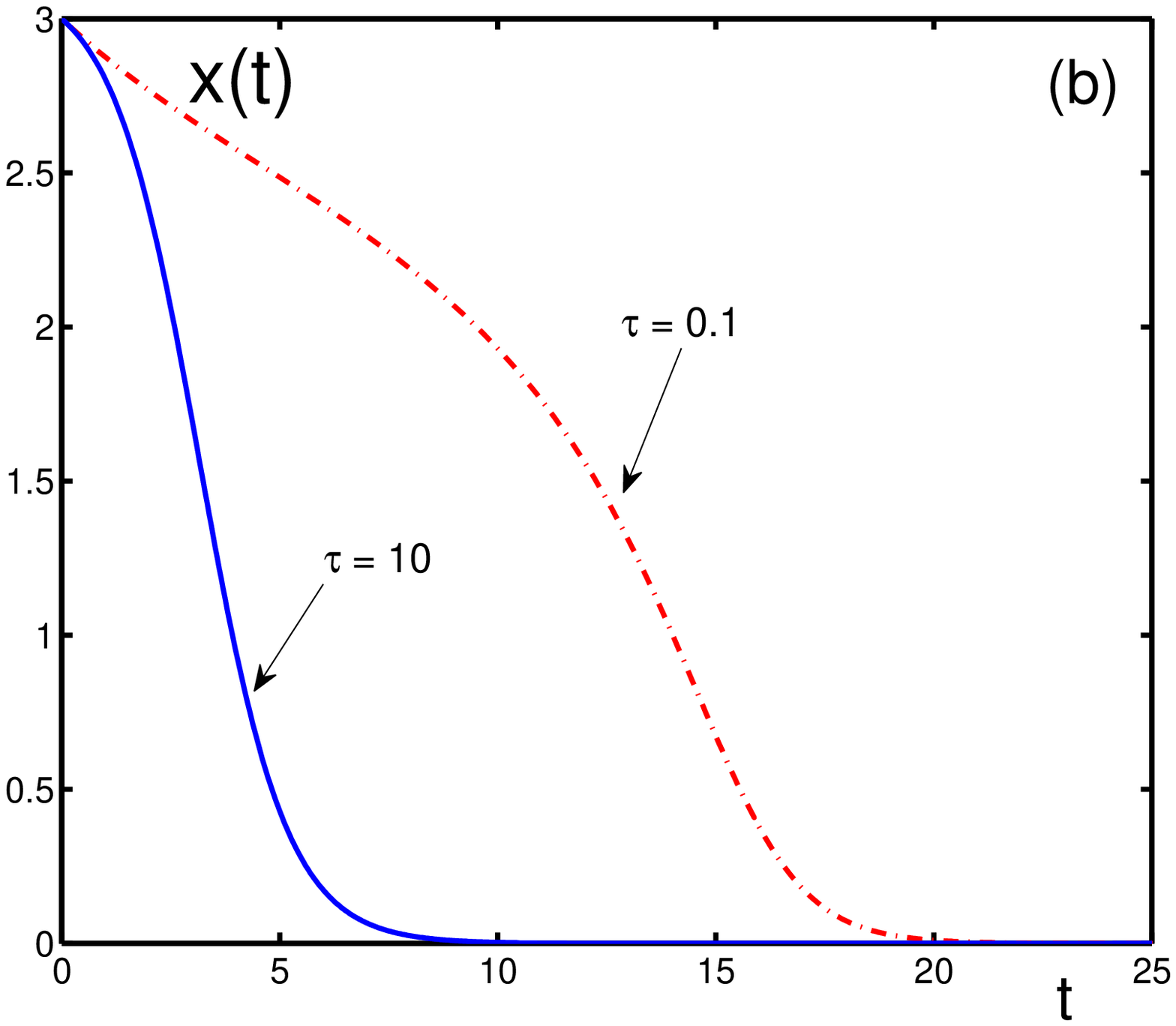} } }
\vspace{9pt}
\centerline{
\hbox{ \includegraphics[width=6.5cm]{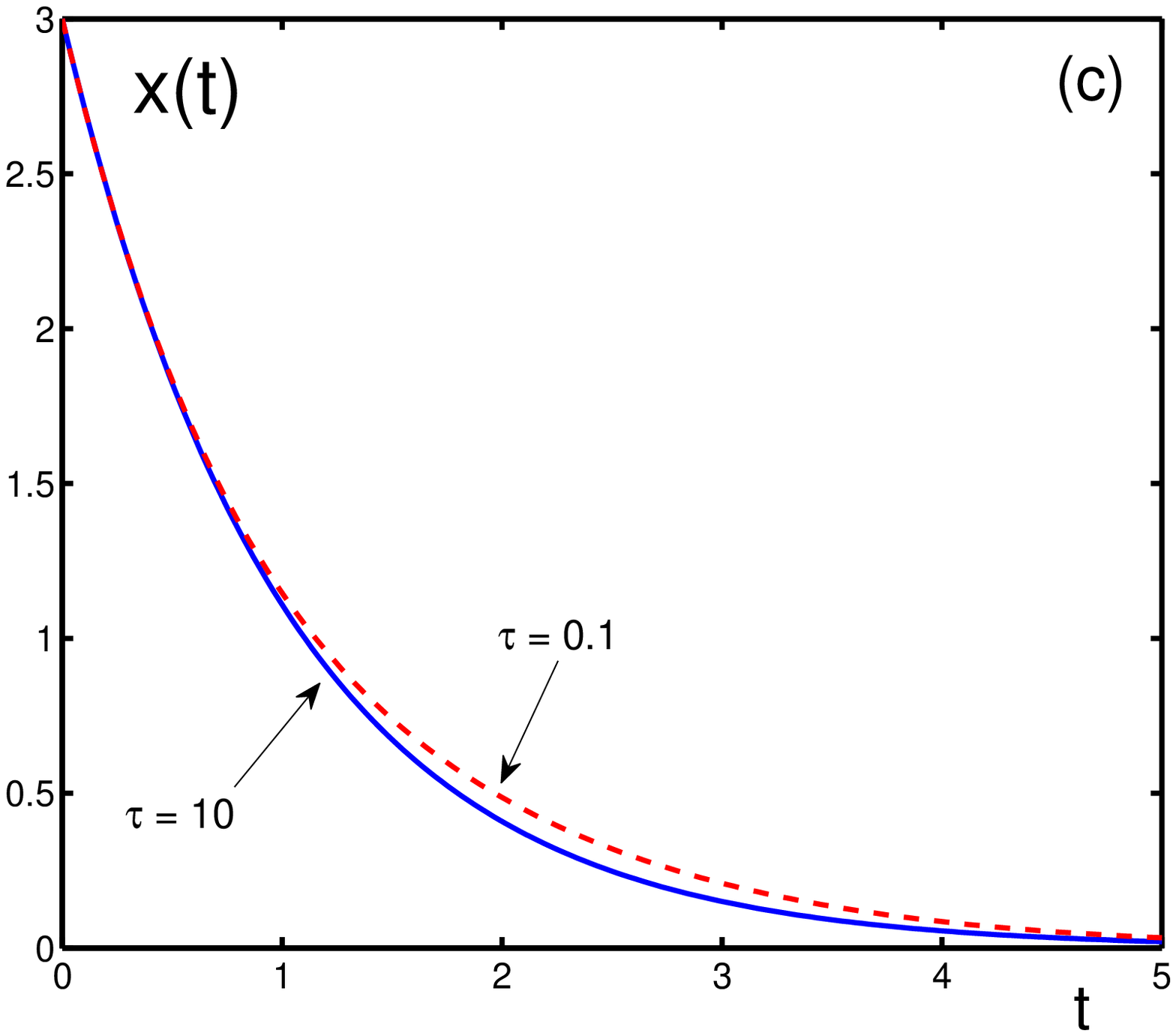} 
}}
\caption{Different types of monotonic convergence to a stationary state, 
under $\sigma_1 = \sigma_2 = -1$, for different parameters: 
(a) $b = 1/e$, $x_0 = 5$, $\tau = 0.7 > \tau^* = 0.627$ (solid line),
$\tau = 0.63 > \tau^* = 0.627$ (dashed line), and $\tau = 0.3 < \tau^* = 0.627$
(dashed-dotted line); (b) $b = 0.38$, $x_0 = 3$, $\tau = 0.1$ (dashed line) and 
$\tau = 10$ (solid line); (c) $b = 2$, $x_0 = 3$, $\tau = 10$ (solid line) and 
$\tau = 0.1$ (dashed line).}
\label{fig:Fig.10}
\end{figure}

\begin{figure}[ht]
\vspace{9pt}
\centerline{
\hbox{ \includegraphics[width=6.5cm]{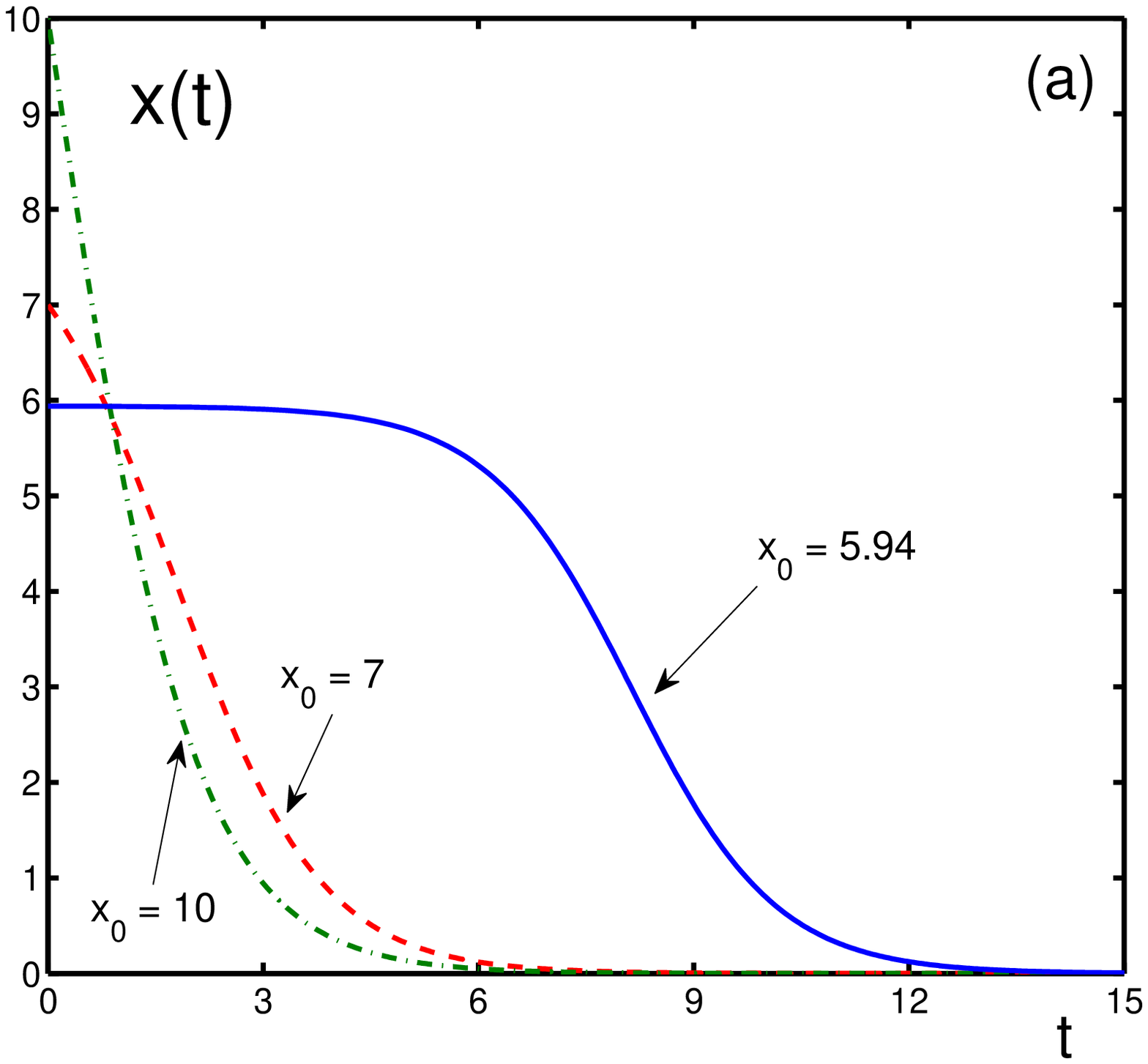} \hspace{2cm}
\includegraphics[width=6.5cm]{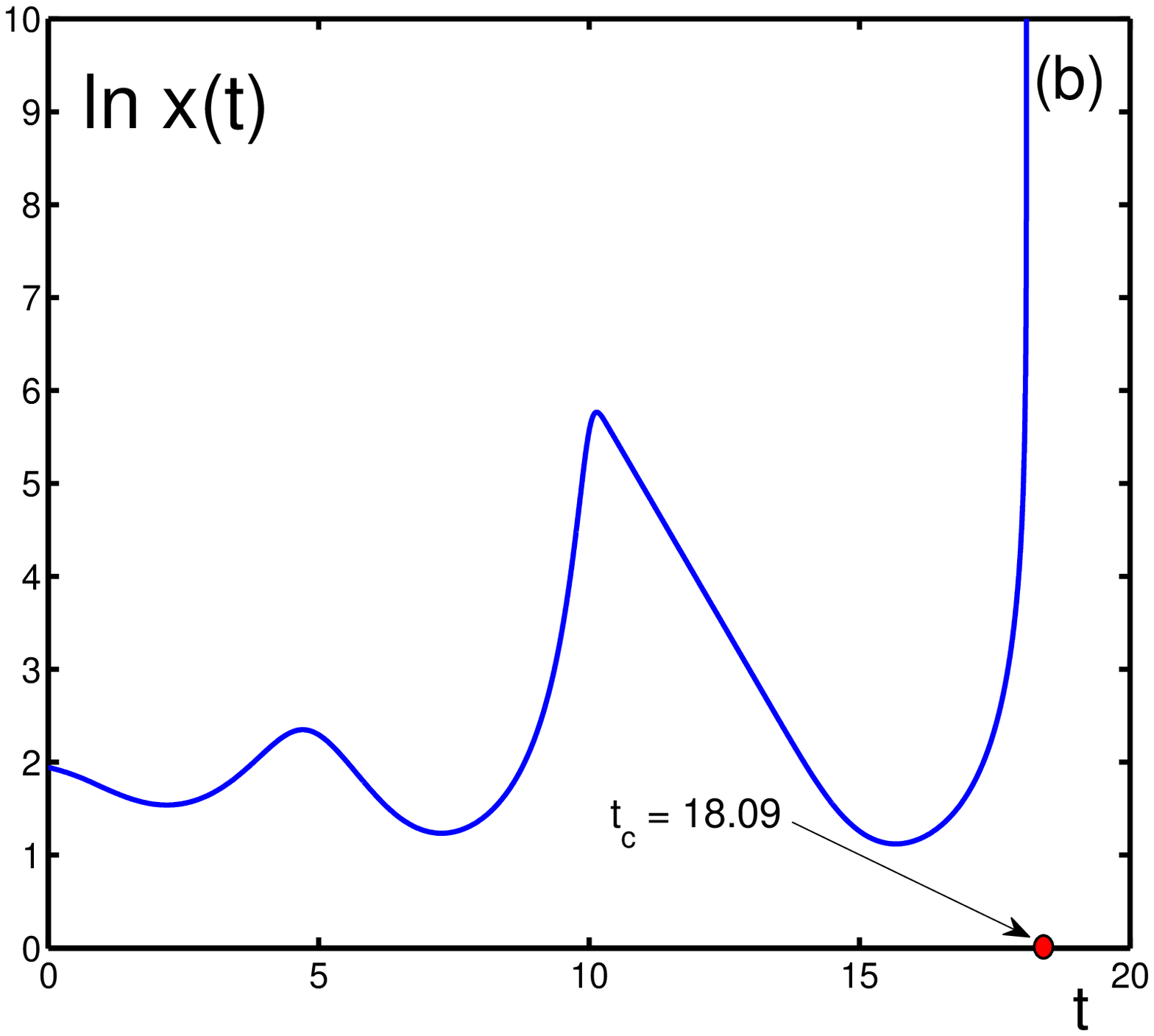} } }
\vspace{9pt}
\centerline{
\hbox{ \includegraphics[width=6.5cm]{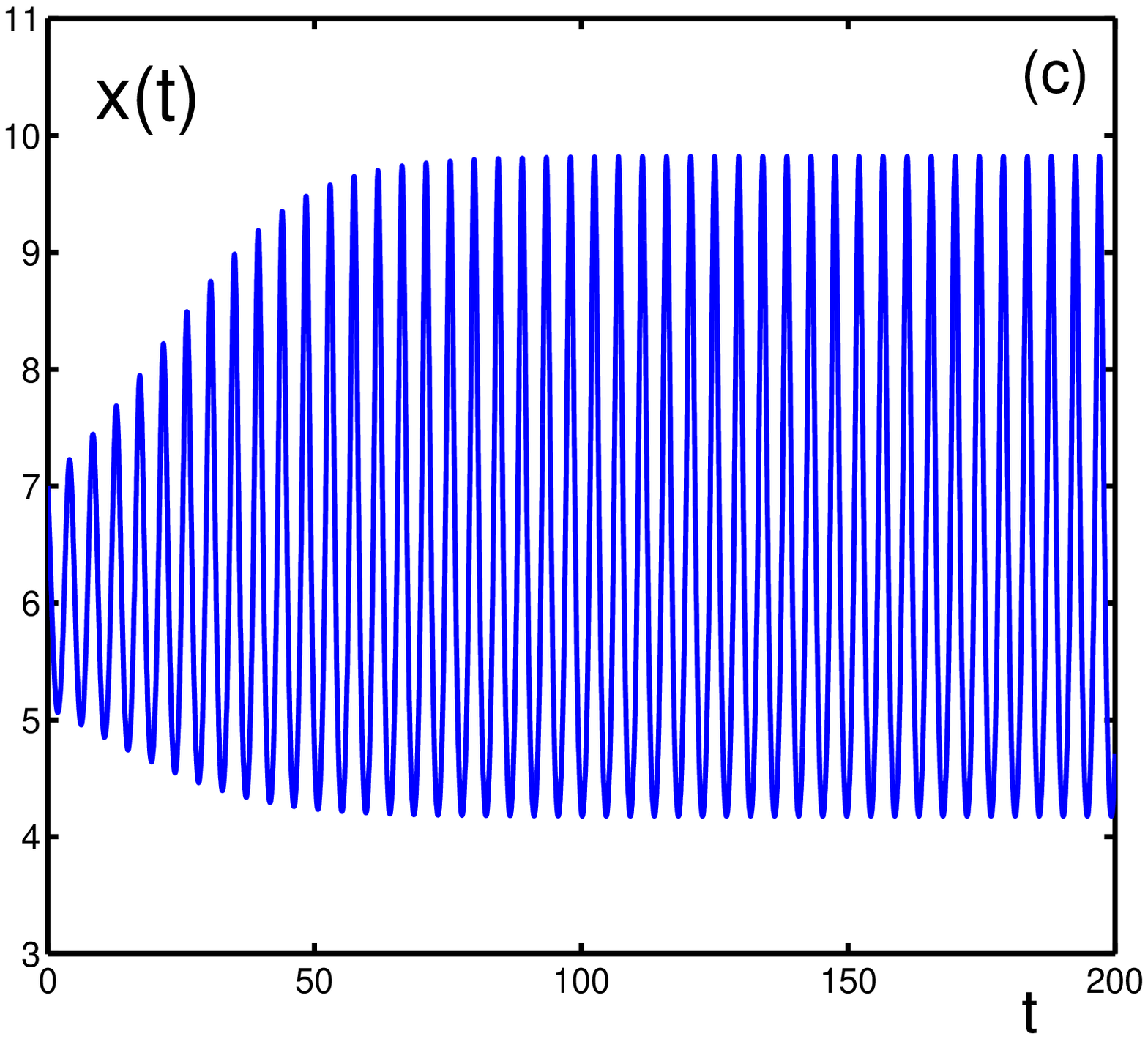} \hspace{2cm}
\includegraphics[width=6.5cm]{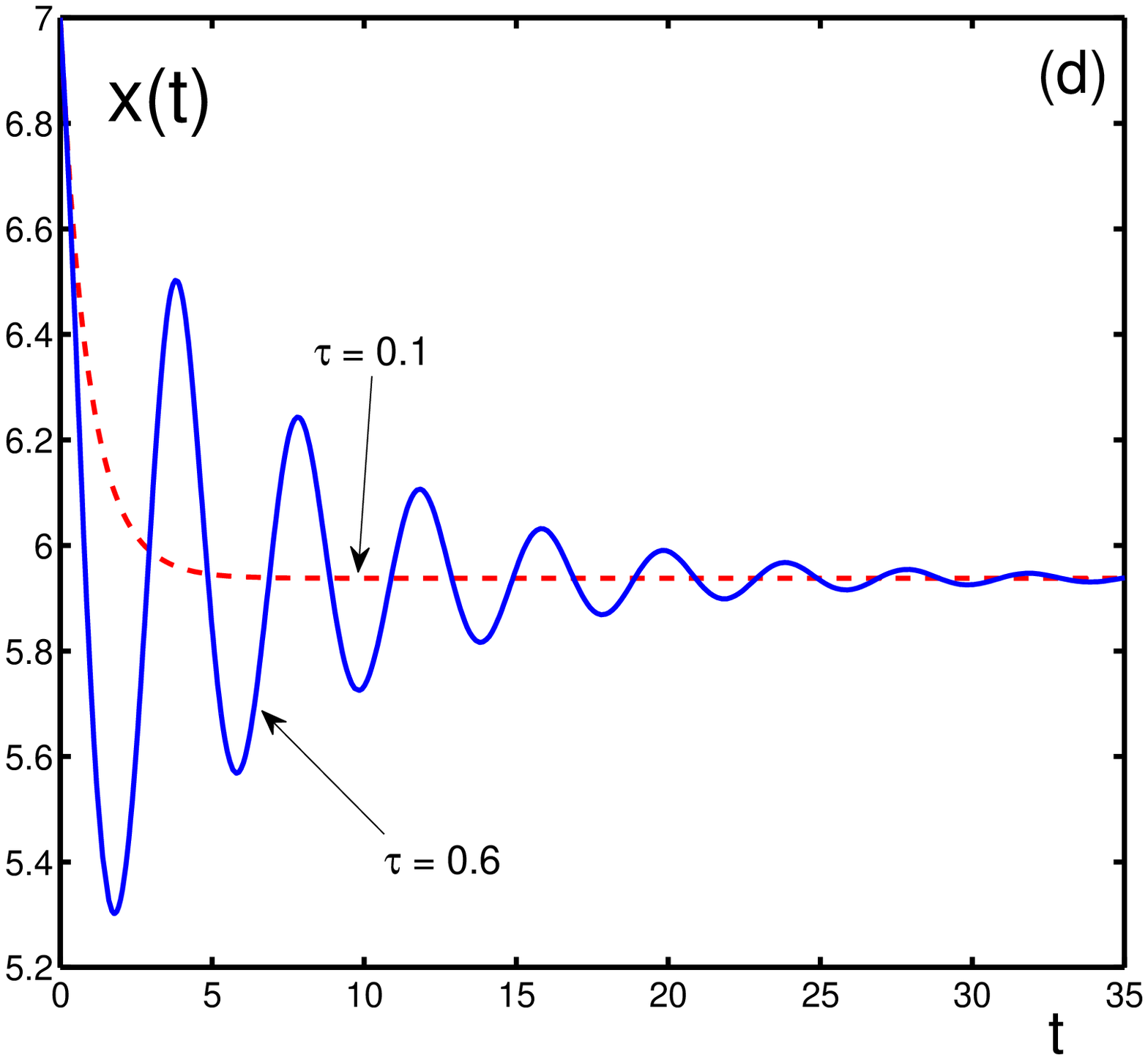} } }
\caption{Change of dynamical regimes, under $\sigma_1 = \sigma_2 = -1$, 
for the history $x_0 > x_3^* = 5.938$, with varying time lags: 
(a) convergence to the stationary state $x_1^* = 0$ for $b = 0.3$, 
$\tau = 20$, $x_0 = 5.94$ (solid line), $x_0=7$ (dashed line), and 
$x_0=10$ (dashed-dotted line); (b) finite-time singularity at 
$t_c = 18.09$ for $b = 0.3$, $x_0 = 7$, $\tau = 0.8$ in the interval 
$\tau_1 < \tau < \tau_2$, with $\tau_1 = 0.777$, $\tau_2 = 1.398$; 
(c) everlasting oscillations for $b=0.3$, $x_0=7$, $\tau =0.68>\tau_3^*=0.661$;  
(d) transformation of the oscillatory convergence to the state $x_3^*=5.938$ 
for $b=0.3$, $x_0=7$ and $\tau=0.6>\tau_3= 0.31$ (solid line) to the monotonic 
decay to the same state for $\tau =0.1<\tau_3 = 0.31$ (dashed line). }
\label{fig:Fig.11}
\end{figure}

\begin{figure}[ht]
\vspace{9pt}
\centerline{
\hbox{ \includegraphics[width=6.5cm]{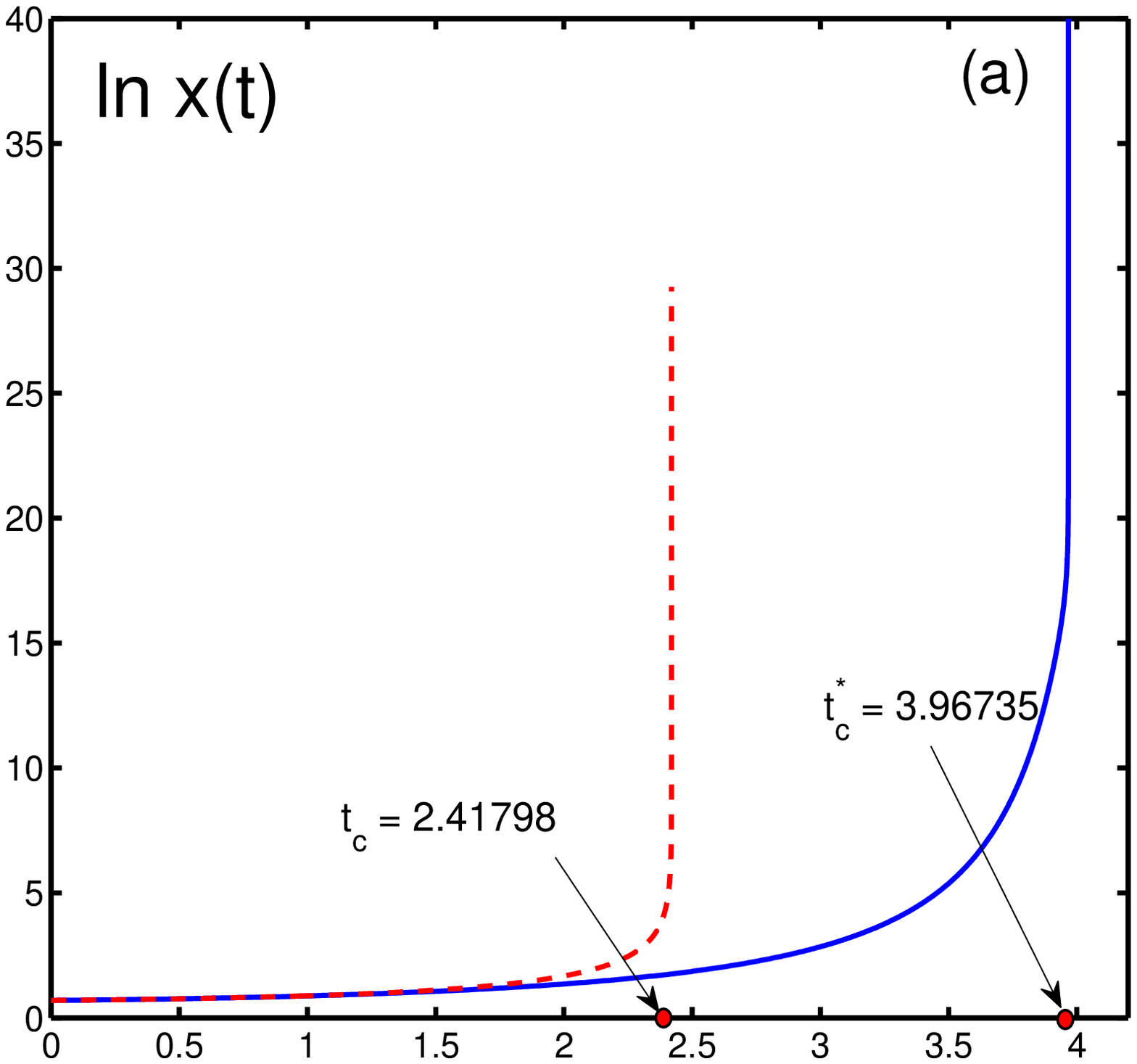} \hspace{2cm}
\includegraphics[width=6.5cm]{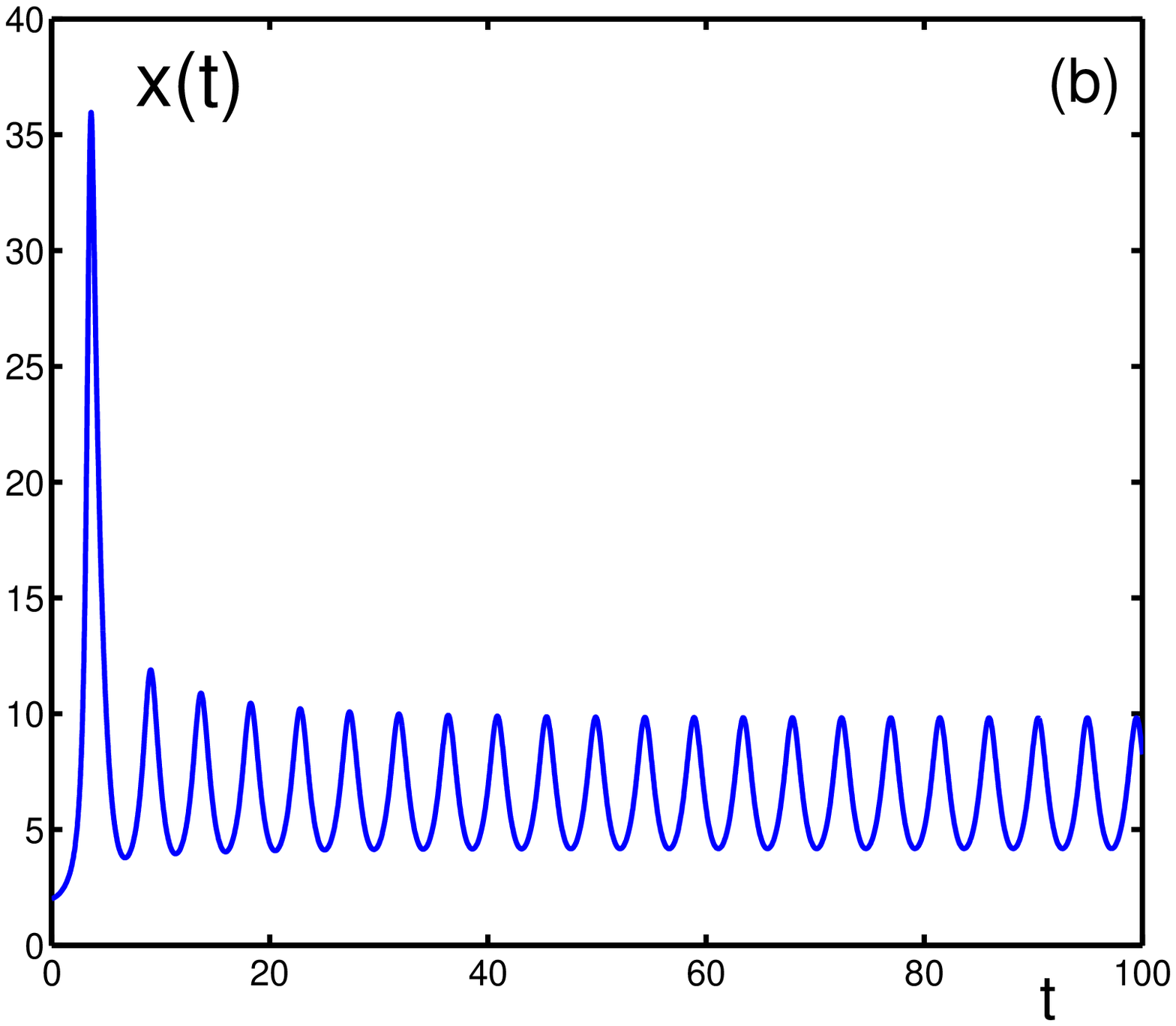} } }
\vspace{9pt}
\centerline{
\hbox{ \includegraphics[width=6.5cm]{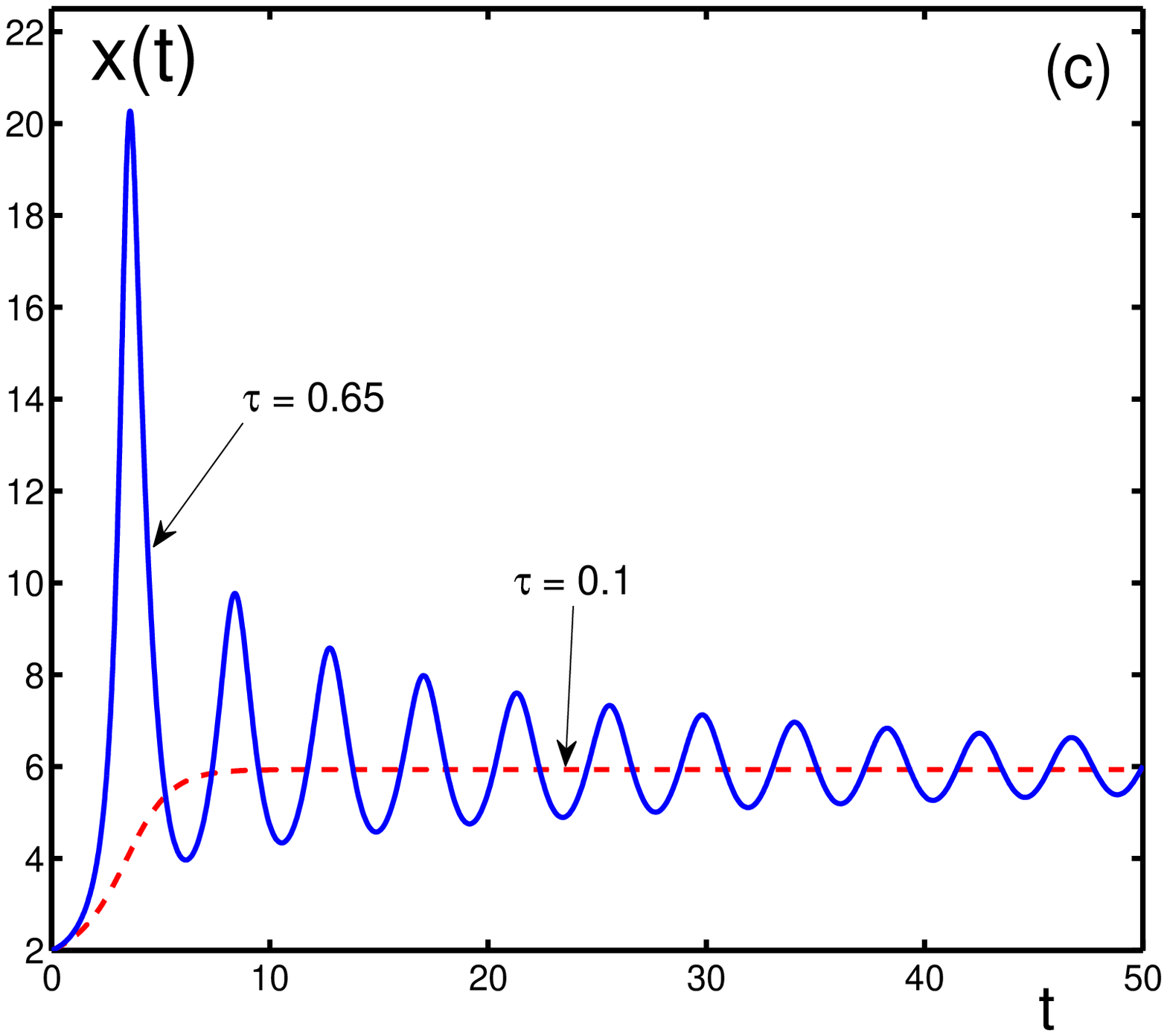} } }
\caption{Dynamical regimes, under $\sigma_1 =\sigma_2=-1$, for the 
history $x_2^*<x_0=2< x_3^*$, with $x_2^*=1.631$, $x_3^* = 5. 938$, 
the fixed production parameter $b = 0.3$, varying only the time lag: 
(a) finite-time singularity at $t_c=2.42$ for $\tau=10>\tau_2=0.713277$ 
(dashed line) and $\tau = 0.713278 > \tau_2$, with the singularity at 
$t_c = 3.967$ (solid line); 
(b) everlasting oscillations for $\tau_3^* < \tau = 0.68 < \tau_2$, 
with $\tau_3^* = 0.661$, $\tau_2 = 0.713277$; (c) convergence to the 
stationary state $x_3^*$ for $\tau_1 < \tau = 0.65 < \tau_3^*$, with 
$\tau_1 = 0.285$ (solid line) and $\tau = 0.1 < \tau_1$ (dashed line). }
\label{fig:Fig.12}
\end{figure}

\begin{figure}[ht]
\vspace{9pt}
\centerline{
\hbox{ \includegraphics[width=6.5cm]{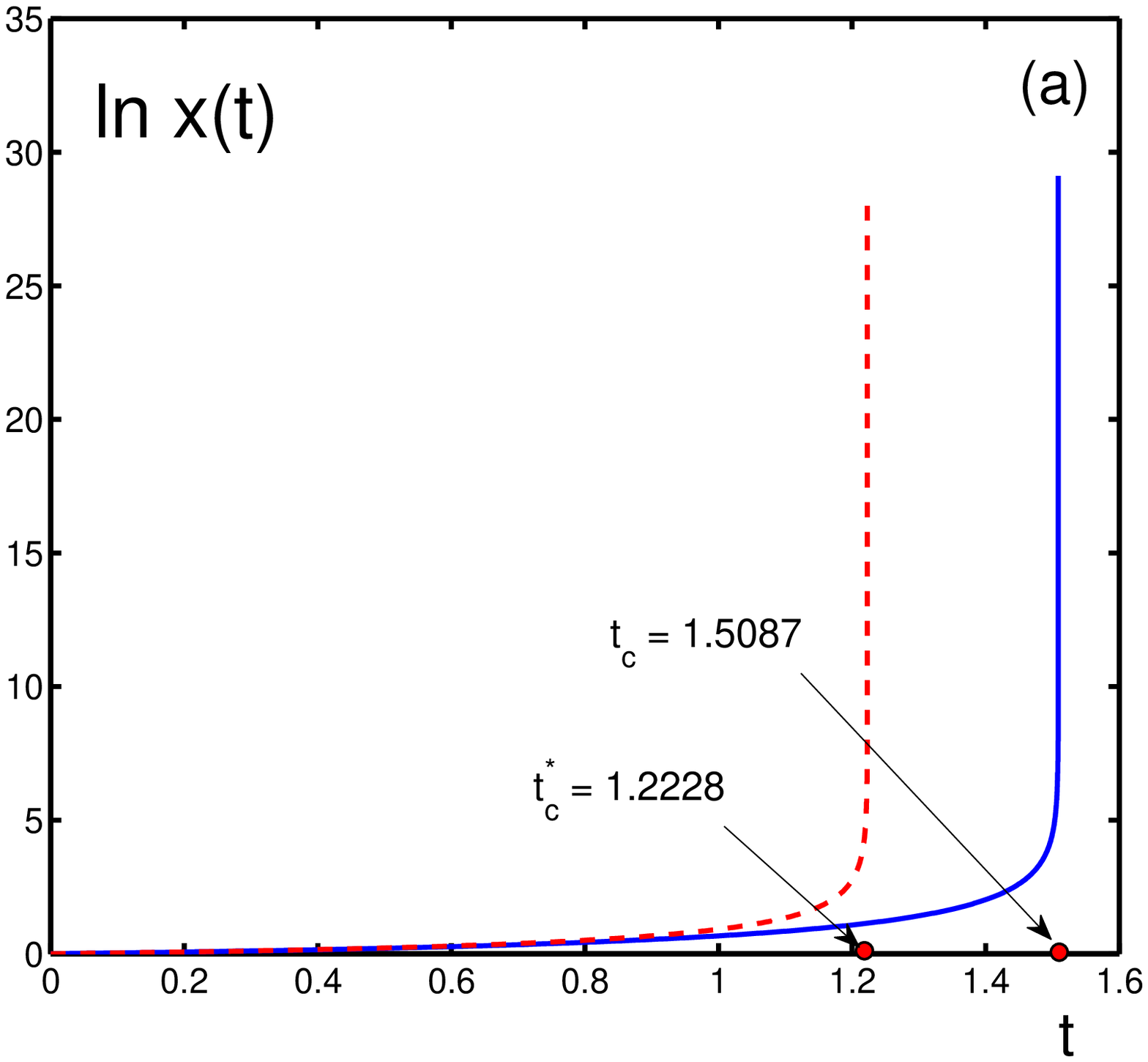} \hspace{2cm}
\includegraphics[width=6.5cm]{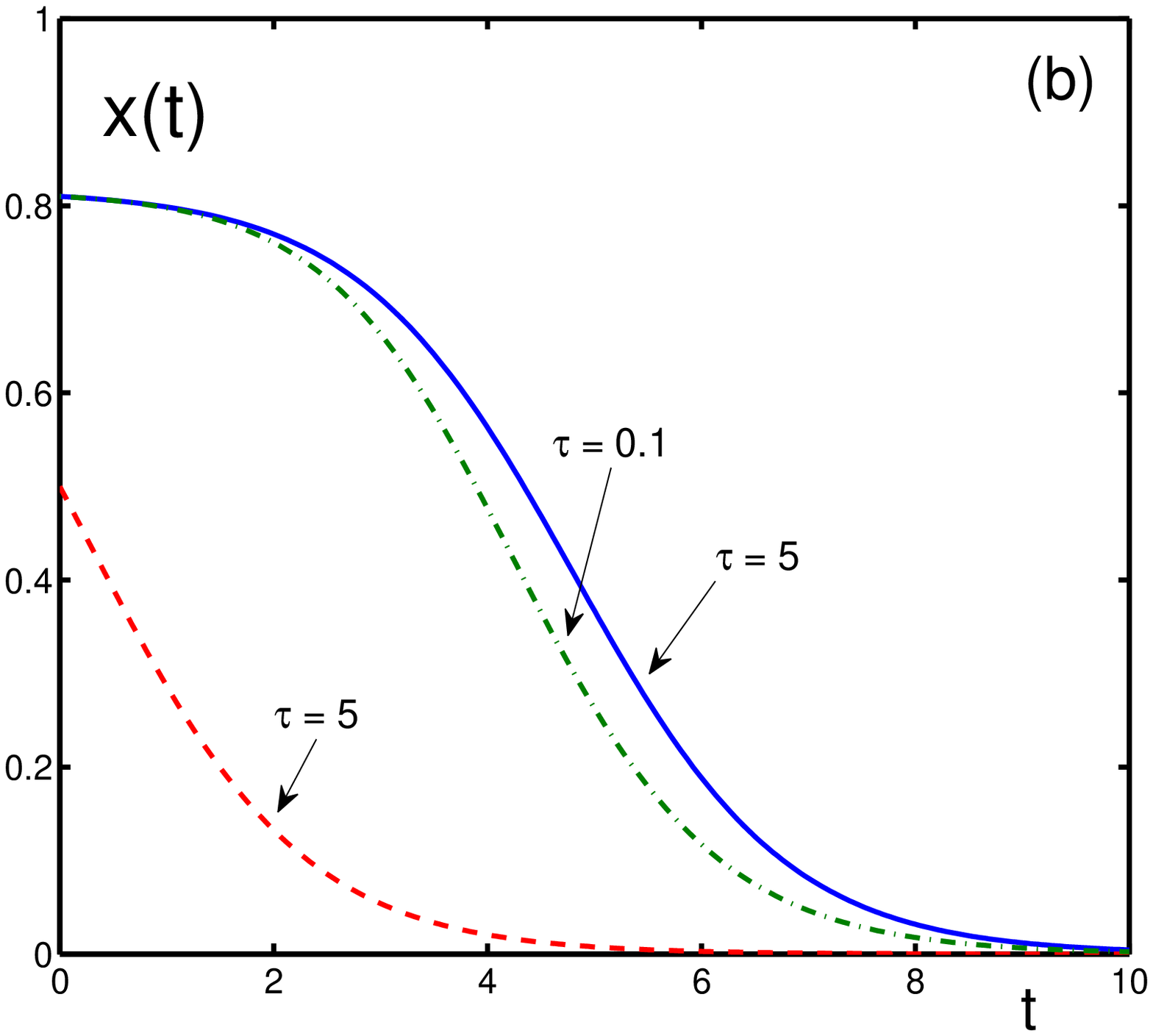} } }
\caption{Dependence of dynamical regimes, under $\sigma_1=\sigma_2=-1$, 
in the case of the destruction parameter $b=-0.25$, on the history and time 
lags: (a) finite-time singularity at $t_c=1.509$ for $x_0=1> x_2^* = 0.816$, 
$\tau = 10 > t_c$ (solid line) and the singularity at $t_c = 1.223$ for 
$\tau=0.1<t_c$ ( dashed line); (b) monotonic degradation to $x_1^*=0$ for the 
same $b = -0.25$, but for $x_0 = 0.81 < x_2^*$, $\tau = 5$ (solid line), for 
$x_0 = 0.5 < x_2^*$, $\tau = 0.1$ (dashed line), and $x_0 = 0.81 < x_2^*$, 
$\tau = 0.1$ (dashed-dotted line).  }
\label{fig:Fig.13}
\end{figure}

\begin{sidewaysfigure}
\centerline{\includegraphics[width=18cm]{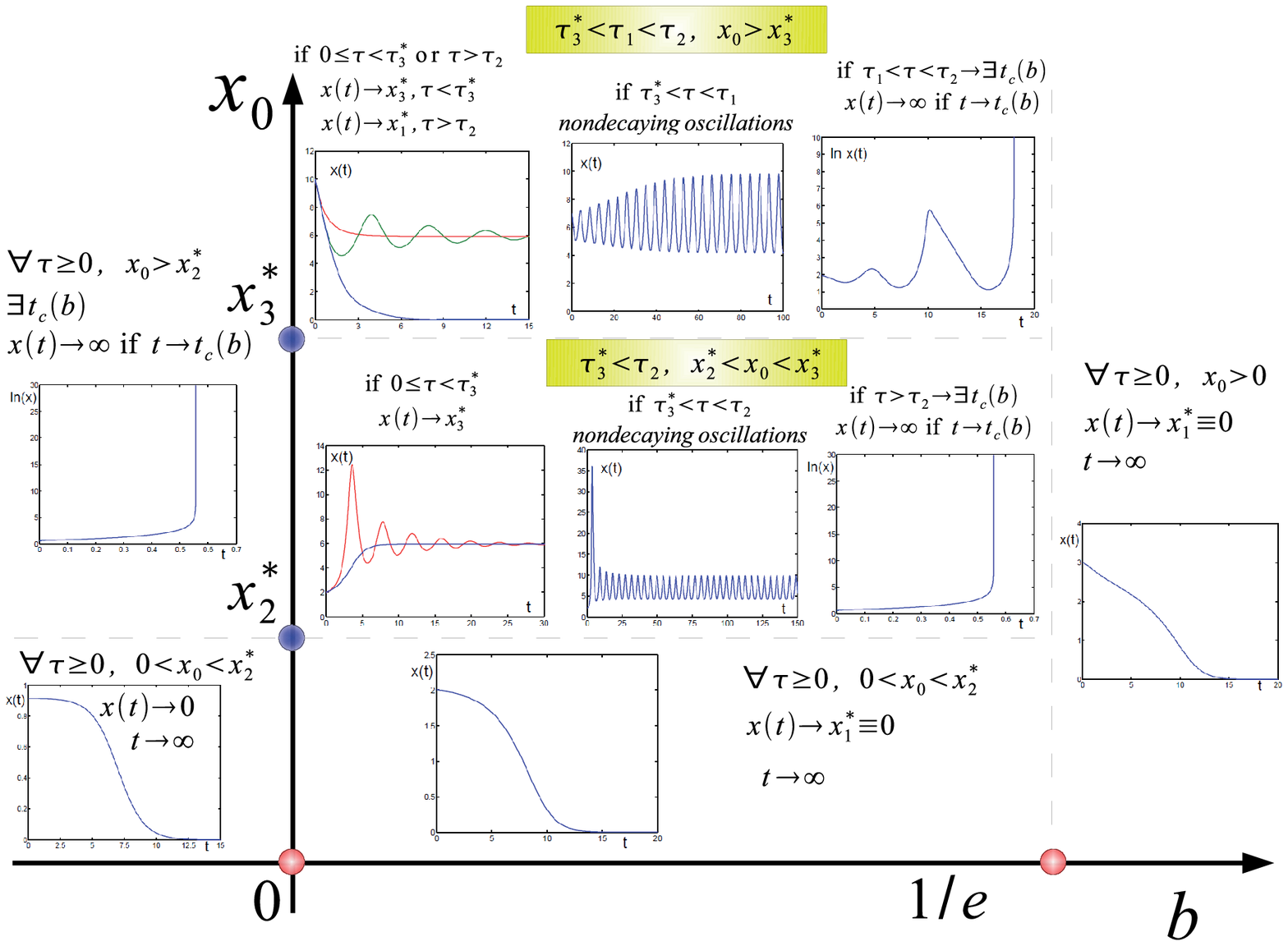}}
\caption{Classification of all possible dynamical regimes, under 
$\sigma_1 = \sigma_2 = -1$, depending on the production parameter $b$, 
time lag $\tau$, and history $x_0$.  
} 
\label{fig:Fig.14}
\end{sidewaysfigure}

\newpage

\section{Dynamics under loss and competition or gain and cooperation 
($\sgm_1\sgm_2 = -1$)}

Contrary to the previous cases, where $\sigma_1$ and $\sigma_2$ were equal, now they
are of opposite signs, so that $\sgm_1\sgm_2 = -1$.

\subsection{Decay under loss and competition}

In such an unfortunate situation, when
\be
\label{33}
 \sgm_1 = -1 \; , \qquad \sgm_2 = 1 \;  ,
\ee
there is only the stationary state $x_1^* = 0$ that is stable for all production
parameters $b \in (- \infty, \infty)$, any time lag $\tau \geq 0$, and arbitrary
history $x_0 \geq 0$. The solutions always monotonically decay to zero, as shown
in Fig. 15.

\begin{figure}[ht]
\centerline{\includegraphics[width=6.5cm]{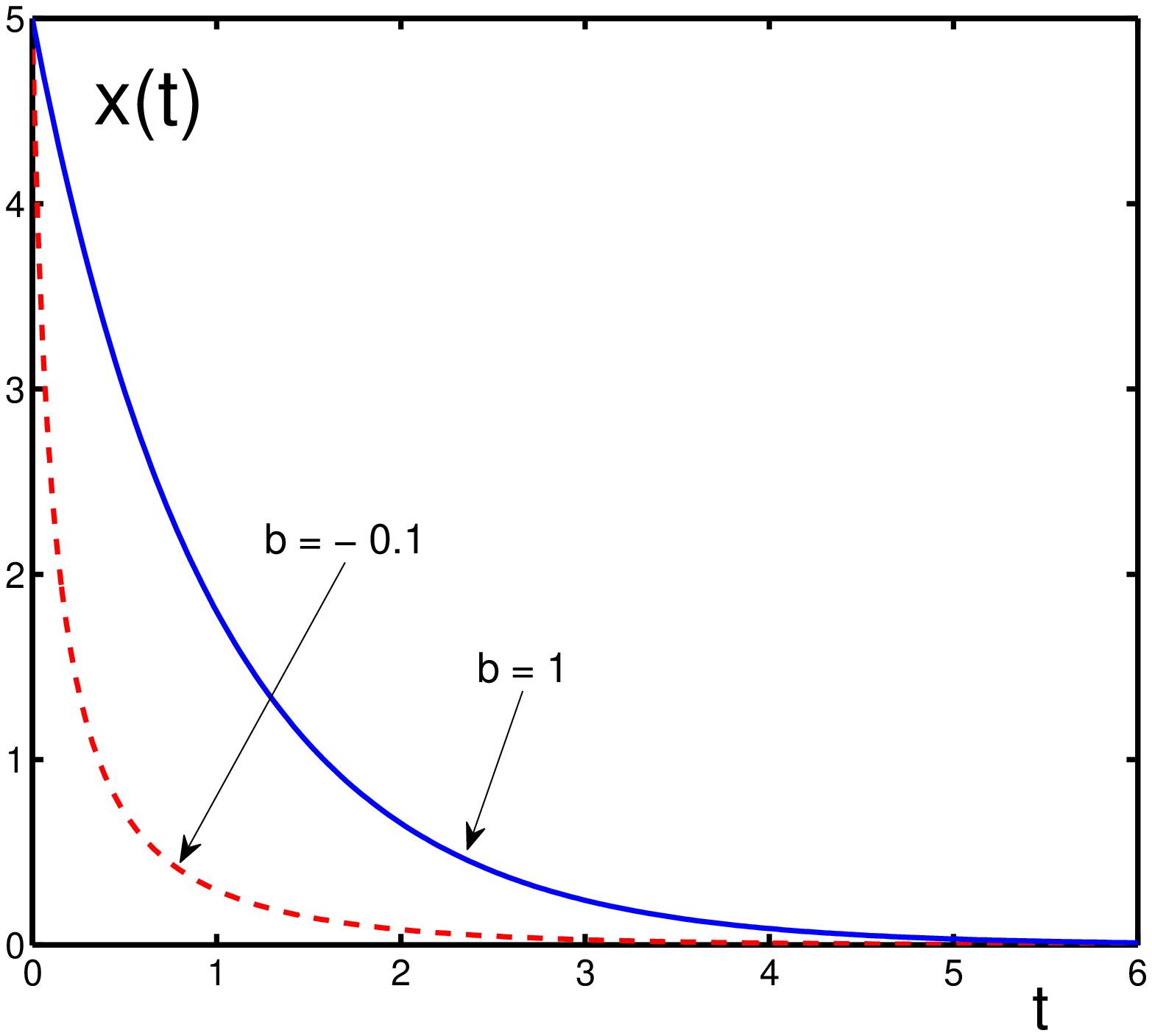}}
\caption{Monotonic decay to zero, under loss and competition 
($\sigma_1 = -1$ and $\sigma_2 = 1$), with the history $x_0 = 5$ and time 
lag $\tau = 10$, for $b = 1$ (solid line) and $b = -0.1$ (dashed line).  } 
\label{fig:Fig.15}
\end{figure}

\subsection{Finite-time singularity or unbounded growth under gain and cooperation
($\sgm_1 = 1, \sgm_2 = - 1$)}

When $\sgm_1 = 1, \sgm_2 = - 1$, the solutions always grow with time, exhibiting
either unbound increase as $t \ra \infty$, or a finite-time singularity at a 
critical time $t_c^*$.

If $b < 0$, the solutions exhibit finite-time singularities. For $\tau\geq t_c$,
where
\be
\label{34}
 t_c = \ln \left ( 1 + \frac{e^{bx_0}}{x_0} \right ) \;  ,
\ee
the point of singularity is $t_c^* = t_c(x_0,b)$. But if $\tau < t_c$, then 
the singularity point, defined numerically, is $t_c^*(x_0,b,\tau) \geq t_c$, 
such that $t_c^* \ra t_c + 0$, as $\tau \ra t_c - 0$. The corresponding 
finite-time singularities are illustrated in Fig. 16a.

When $b > 0$, there can occur either unbounded growth as $t \ra \infty$ or a
finite-time singularity. If $0 \leq \tau < \tau^*$, where $\tau^* = \tau^*(x_0,b)$
is defined numerically, the solution unboundedly grows as $t \ra \infty$. For
$\tau^* \leq \tau < t_c$, there exists a critical time $t_c^* = t_c^*(\tau) > t_c$
at which the solution diverges. And when $\tau \geq t_c$, the divergence happens 
at $t_c^* = t_c$, where $t_c$ is given by expression (\ref{38}). The change of 
the regime for the same production parameter $b$ and history $x_0$, but a varying 
time lag $\tau$ is illustrated in Fig. 16b.

\begin{figure}[ht]
\vspace{9pt}
\centerline{
\hbox{ \includegraphics[width=6.5cm]{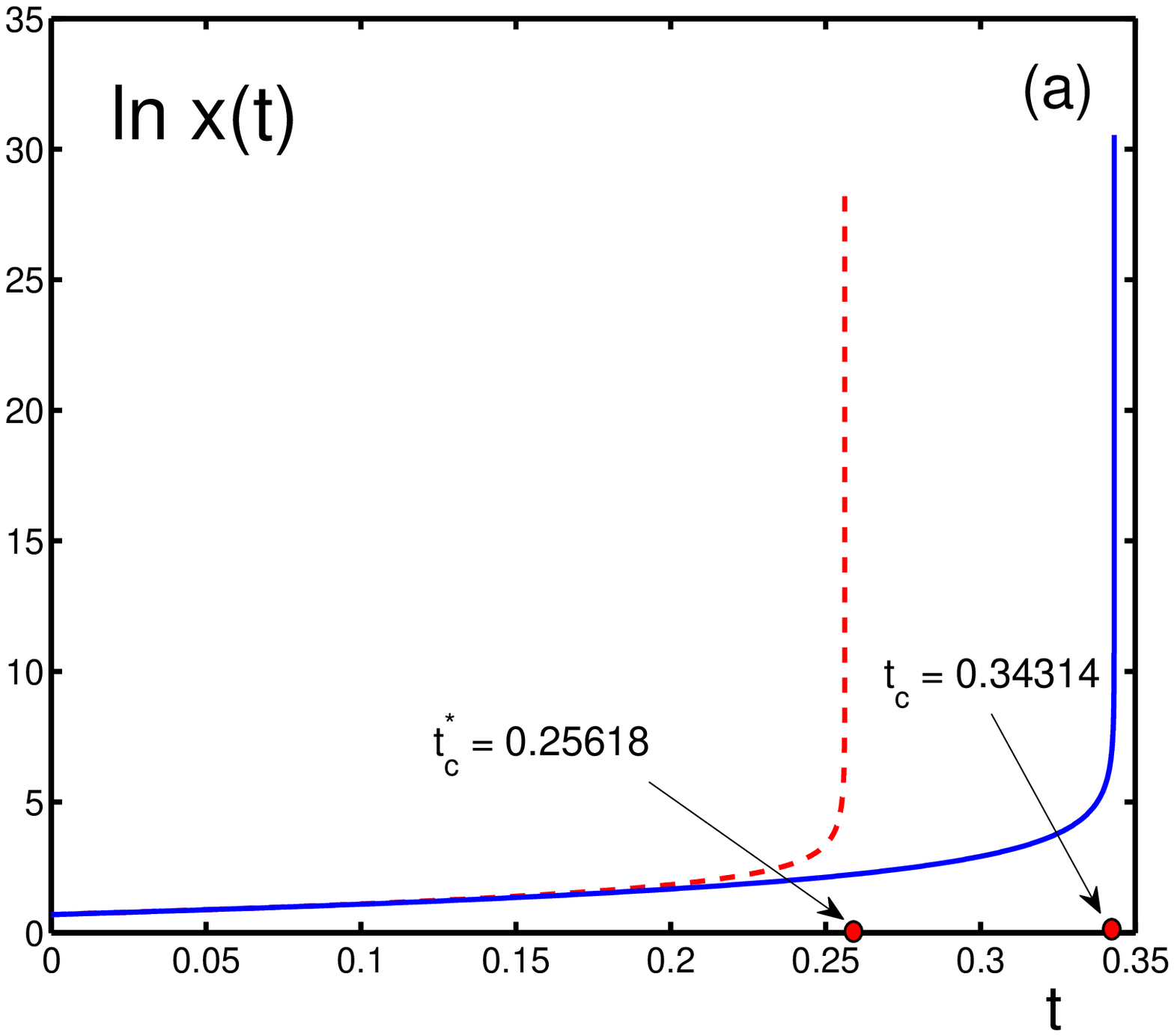} \hspace{2cm}
\includegraphics[width=6.5cm]{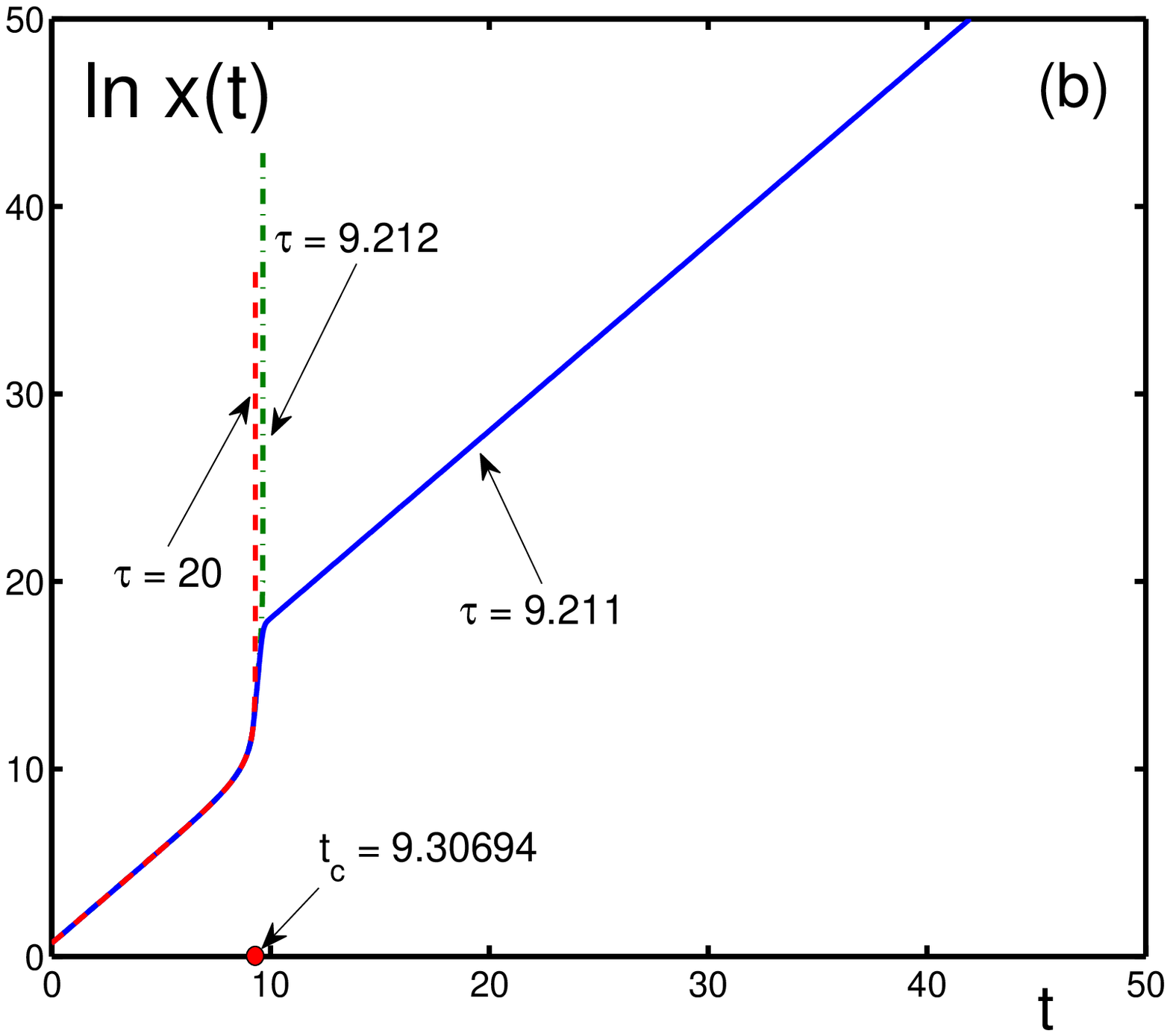} } }
\caption{{\bf Fig. 16}. Behavior of the solutions in logarithmic scale, under $\sigma_1 = 1$, 
$\sigma_2 = -1$, exhibiting either finite-time singularities or unbounded 
growth at $t \ra \infty$: (a) finite-time singularity, with the divergence 
at $t_c = 0.343$ for $b = -0.1$, $x_0 = 2$, $\tau = 10 > t_c$ (solid line) 
and with the divergence at $t_c^* = 0.256$ for $b = -0.1$, $x_0 = 2$, 
$\tau = 0.01 < t_c$ (dashed line); (b) the change of regime for $b = 5$, 
$x_0=2$ and varying time lag, when the finite-time singularity at $t_c=9.307$ 
for $\tau = 20 > t_c$ (dashed line), or at $t_c^* = 9.655 > t_c$ for 
$\tau^* < \tau = 9.212 < t_c$, where $\tau^* = 9.2117$ (dashed-dotted line), 
changes to the unbounded growth as $t\ra\infty$ for $\tau=9.211<\tau^*$ 
(solid line).}
\label{fig:Fig.16}
\end{figure}

\section{Influence of noise}

\subsection{Stochastic differential equation}

In the presence of noise, the evolution equation (7) becomes the stochastic
differential equation
\be
\label{35}
 dx = g(x,t) dt + \al d W_t \;  ,
\ee
where $\alpha = \sqrt{2 D}$ characterizes the noise strength, with $D$ being
the diffusion coefficient, and where
\be
\label{36}
  g(x,t) = \sgm_1 x - \sgm_2 x^2 \exp\{ - bx(t-\tau) \} \; .
\ee
We consider Eq. (\ref{35}) in the sense of Ito, where $W_t$ is the standard 
Wiener process.

The addition of the noise does not influence much those solutions that do not
exhibit finite-time singularities, while the latter can be strongly influenced
by even weak noise. Therefore we concentrate our attention on the most 
interesting case of the noise influence on the solutions with finite-time 
singularities.

\subsection{Influence of noise on finite-time singularities}

Recall that a finite-time singularity can occur only in the case of cooperation,
when $\sigma_2 = -1$. Even rather weak noise can essentially shift the 
singularity point. Moreover the same noise strength, in different stochastic 
realizations, shifts the singularity point in a random way. Under the occurrence 
of a finite-time singularity, the influence of noise turns out to be more 
important than the variation of the time lag. This is in agreement with the Mao 
theorem \cite{Mao_31}, according to which there can exist a finite range of time 
lags for which the solution to the differential delay equation is close to that 
of the related ordinary differential equation. Figure 17 illustrates the 
influence of noise on the singularity point.

\begin{figure}[ht]
\vspace{9pt}
\centerline{
\hbox{ \includegraphics[width=6.5cm]{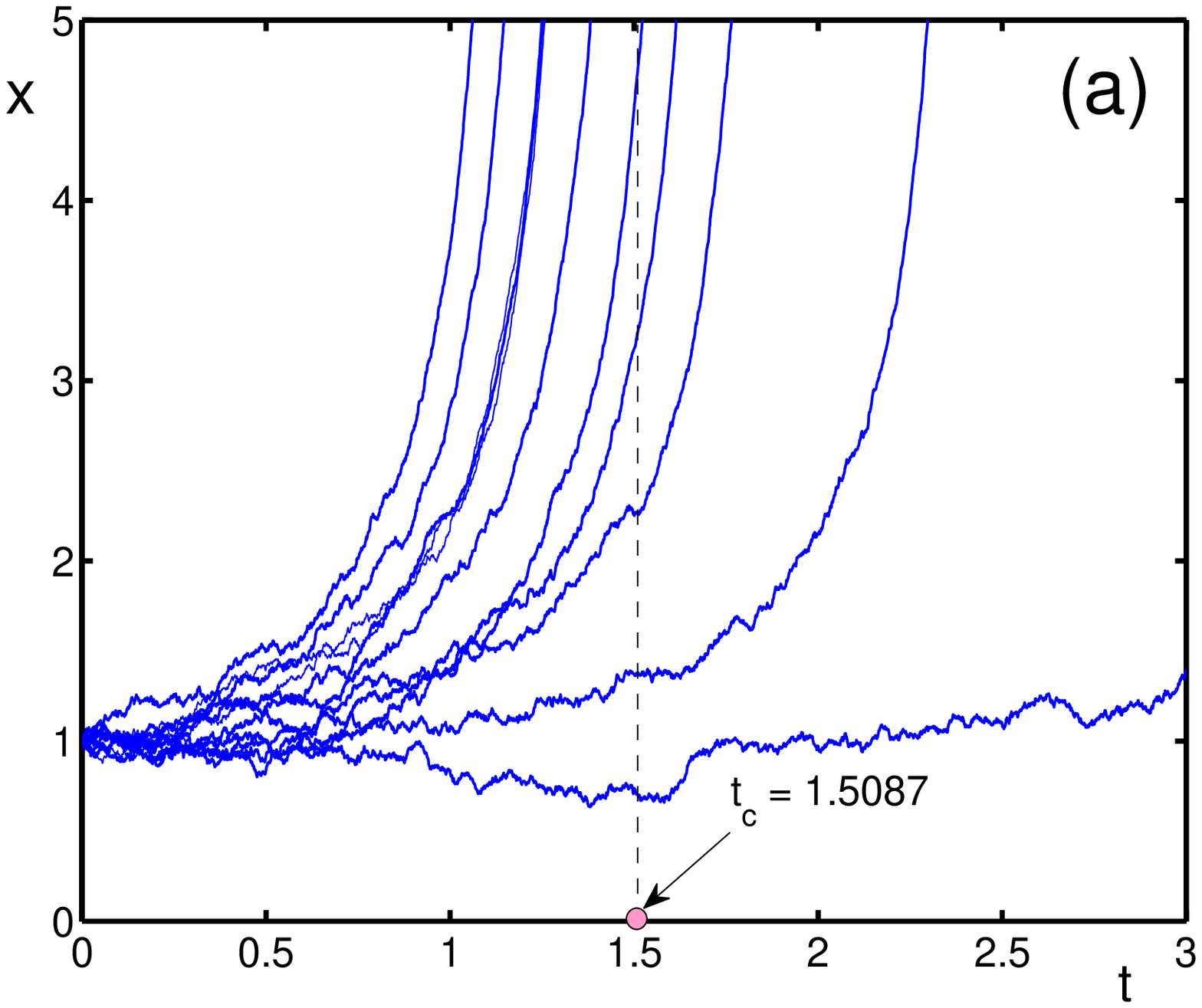} \hspace{2cm}
\includegraphics[width=6.5cm]{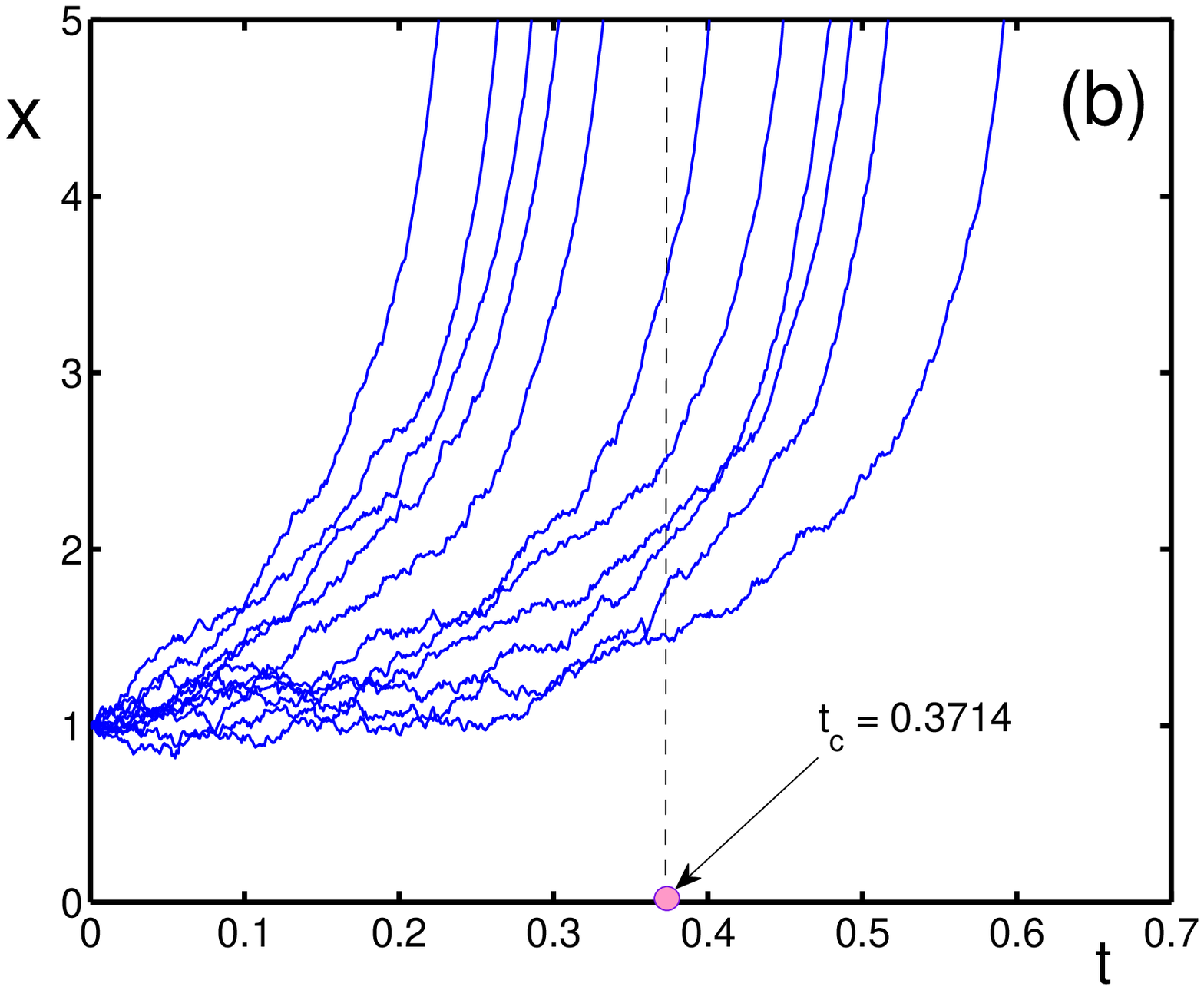} } }
\vspace{9pt}
\centerline{
\hbox{ \includegraphics[width=6.5cm]{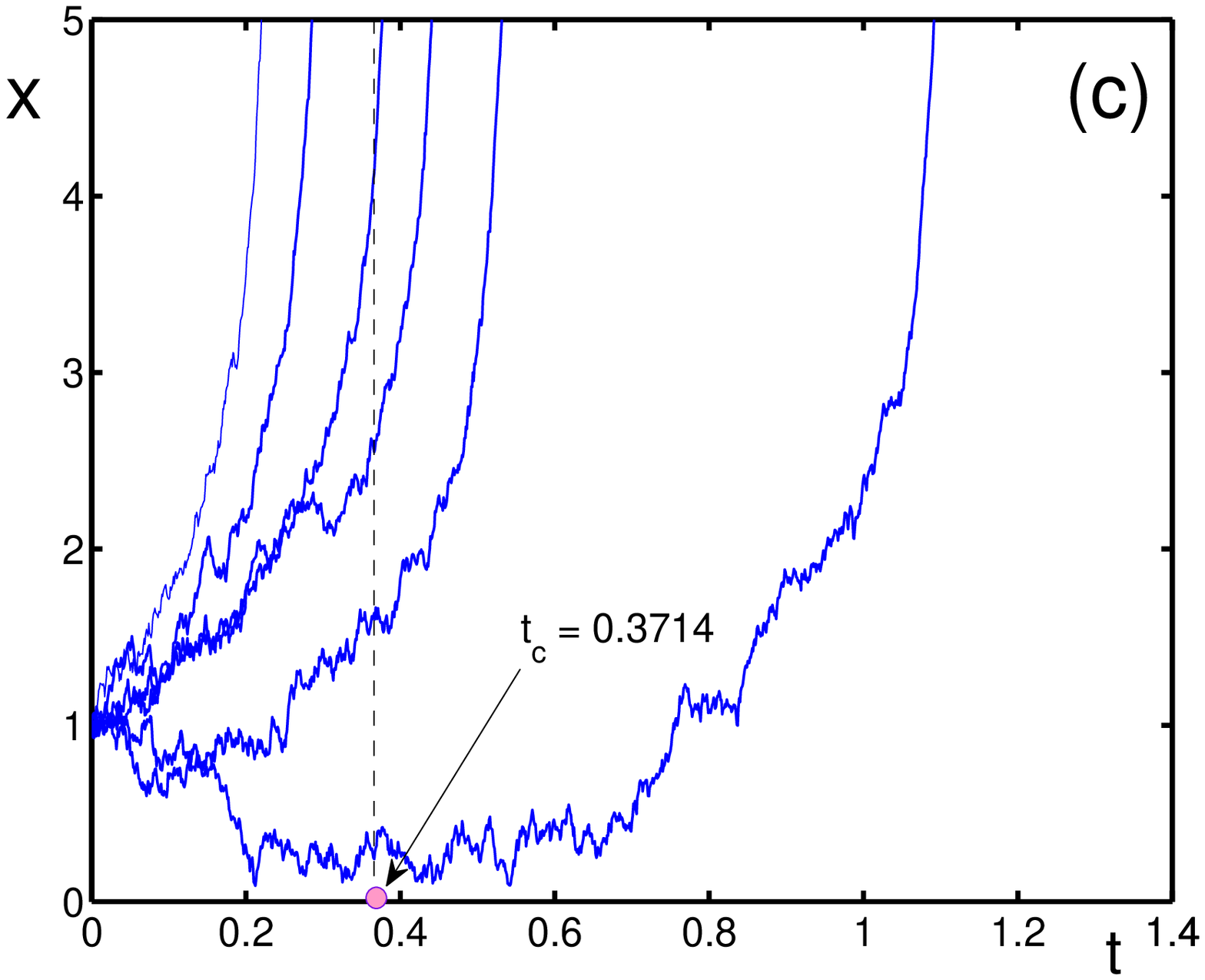} \hspace{2cm}
\includegraphics[width=6.5cm]{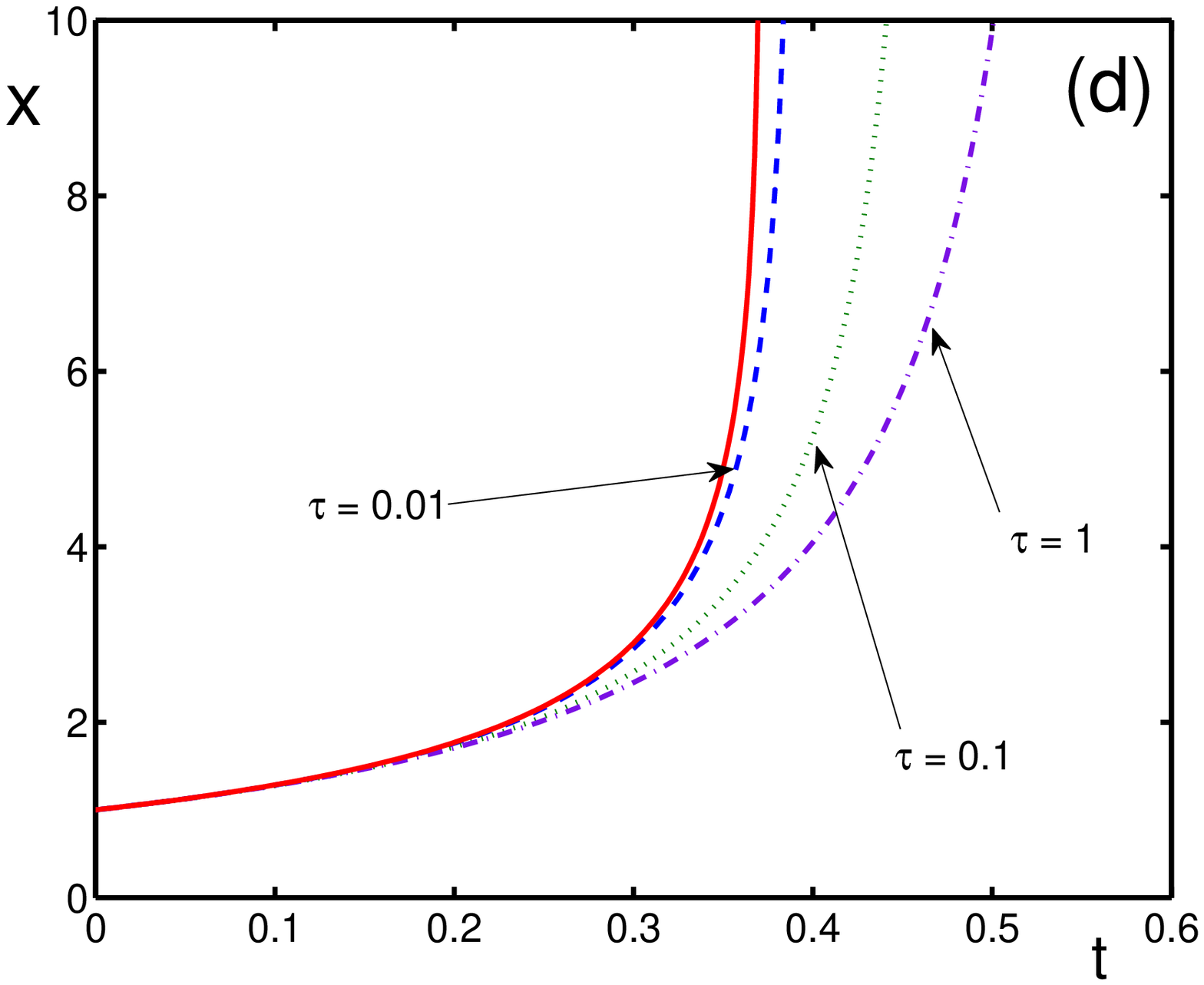} } }
\caption{Influence of noise, in the case of $b=-0.25$ and $x_0=1$, on the 
point of the finite-time singularity: (a) several realizations of stochastic 
trajectories, under $\sigma_1 = \sigma_2 = -1$ and $\tau > t_c = 1.509$ for 
the same noise strength $\alpha = 0.25$; (b) stochastic trajectories, under 
$\sigma_1 = 1$, $\sigma_2 = -1$, with $\tau \ll t_c = 0.371$, for the same 
noise strength $\al = 0.5$; (c) stochastic trajectories, with the parameters 
as in (b), but for the larger noise strength $\al=1$; (d) singular solutions
for the parameters as in (b), but with $\alpha = 0$, for different time lags,
$\tau=0$ (solid line), $\tau=0.01$ (dashed line), $\tau=0.1$ (dotted line), 
and $\tau = 1$ (dashed-dotted line).  }
\label{fig:Fig.17}
\end{figure}

\subsection{Fokker-Planck equation and condition for existence of a stationary
probability distribution}

The stochastic differential equation (\ref{35}) corresponds to the Fokker-Planck 
equation
\be
\label{37}
 \frac{\prt}{\prt t} \; P(x,t) = -\; \frac{\prt}{\prt x} [ g(x,t) P(x,t) ] +
D\; \frac{\prt^2}{\prt x^2}\; P(x,t)
\ee
for the distribution function $P(x,t)$, with $D = \alpha^2/2$. Looking for a 
solution that would be defined for all $x \in [0,\infty)$ and all $t \geq 0$ 
requires the existence of a stationary distribution
\be
\label{38}
 P(x) \equiv \lim_{t\ra\infty} P(x,t) \;   .
\ee
The latter is defined by the equation
\be
\label{39}
 D\; \frac{\prt^2}{\prt x^2}\; P(x) - \;\frac{\prt}{\prt x} [ g(x) P(x) ] = 0 \; ,
\ee
in which
\be
\label{40}
 g(x) = \sgm_1 x - \sgm_2 x^2 \exp(-bx) \;  .
\ee

Equation (\ref{39}) has to be complemented by boundary conditions, which are 
usually taken in the form of absorbing boundary conditions
\be
\label{41}
 \lim_{x\ra\infty} P(x) = 0 \; , \qquad  \lim_{x\ra\infty} g(x)P(x) = 0 \; ,
\ee
since these conditions allow for the normalization of the distribution as
\be
\label{42}
 \int_0^\infty P(x) \; dx = 1 \; .
\ee

The solution to Eq. (\ref{39}) is
\be
\label{43}
 P(x) = C \exp \{ -\bt U(x) \} \;  ,
\ee
with the normalization constant $C$, the effective temperature
$\beta \equiv 1/D = 2/\alpha^2$ and the effective potential
\be
\label{44}
 U(x) = -\; \frac{\sgm_1}{2}\; x^2  - \;
\frac{\sgm_2}{b^3} \left ( 2 + 2bx + b^2 x^2 \right ) e^{-bx} \;  .
\ee

To satisfy the boundary condition for $P(x)$, it is necessary that
\be
\label{45}
 U(x) \ra \infty \qquad (x\ra \infty) \;  .
\ee
From the asymptotic expression
\be
\label{46}
U(x) \simeq - \left ( \frac{\sgm_1}{2} + \frac{\sgm_2}{b}\; e^{-bx} \right ) x^2
\qquad ( x\ra\infty) \; ,
\ee
we find that condition (\ref{45}) is valid in the following cases.

For positive production parameters, the limit (46) yields
$$
 U(x) \simeq - \; \frac{\sgm_1}{2} \; x^2 \qquad (b > 0, \; x\ra\infty) \; ,
$$
which means that condition (45) is satisfied provided that 
\be
\label{47}
\sgm_1 = -1 \; , \qquad \sgm_2 =\pm 1 \; , \qquad b > 0 \; .
\ee
When the feedback is destructive, then the limit (46) gives
$$
 U(x) \simeq \frac{\sgm_2}{|b|} \;  e^{|b|x} \qquad (b < 0, \; x\ra\infty) \; ,
$$
hence condition (45) implies that
\be
\label{48}
\sgm_1 = \pm 1 \; , \qquad \sgm_2 = 1 \; , \qquad b < 0 \; .
\ee
If $b = 0$, then Eqs. (39) and (40) show that the limit (45) requires the same
conditions (48) as for $b < 0$.  

The stationary distribution (\ref{43}) possesses maxima when the effective 
potential (\ref{44}) displays minima. The latter correspond to the stable fixed 
points $x^*$ of the differential equation for $x(t)$. Combining the analysis of 
the conditions for the existence of the distribution $P(x)$ with the conditions 
for the existence of stable fixed points of the delay differential equation (7), 
we find the following: When the solution $x(t)$, for any history $x_0$ and 
$\tau \ra 0$, converges to a fixed point, then $P(x)$ exists. Conversely, when 
there is, at least for some history $x_0$ and $\tau \ra 0$, an unbound solution 
$x(t)$, diverging either at a finite-time singularity or at increasing time 
$t \ra \infty$, then $P(x)$ does not exist. Summarizing, we come to the 
conclusion.

\vskip 2mm

{\bf Statement}. {\it The necessary and sufficient condition for the existence
of the distribution $P(x)$ is the convergence, for any history $x_0 \geq 0$ and
$\tau \ra 0$, of the solution $x(t)$ to a fixed point}.

\vskip 2mm

{\bf Remark}. Note the importance of the limit $\tau \ra 0$. As follows from 
the analysis of the previous sections, $P(x)$ can exist, though $x(t)$ diverges 
for some finite $\tau > 0$.

\vskip 2mm

Different shapes of the distribution $P(x)$, as a function of $x$, are shown 
for the case of a single fixed point $x_1^* = 0$ in Fig. 18 and for either the 
occurrence of the bistability region, with two fixed points $x_1^* = 0$ and 
$x_3^* > 0$, or for the case of one nontrivial fixed point $x_2^* > 0$ in Fig. 19.

\begin{figure}[ht]
\vspace{9pt}
\centerline{
\hbox{ \includegraphics[width=6.5cm]{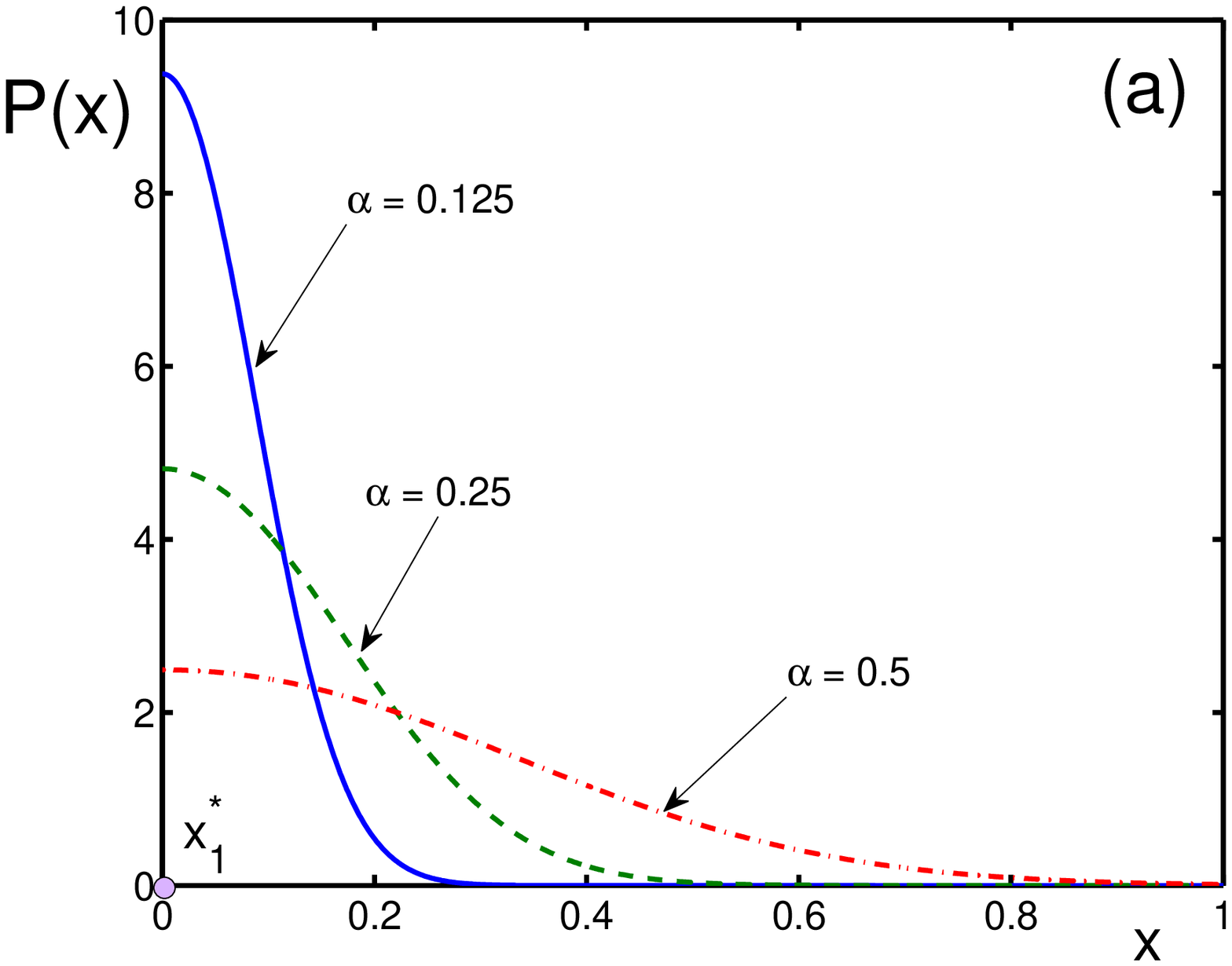} \hspace{2cm}
\includegraphics[width=6.5cm]{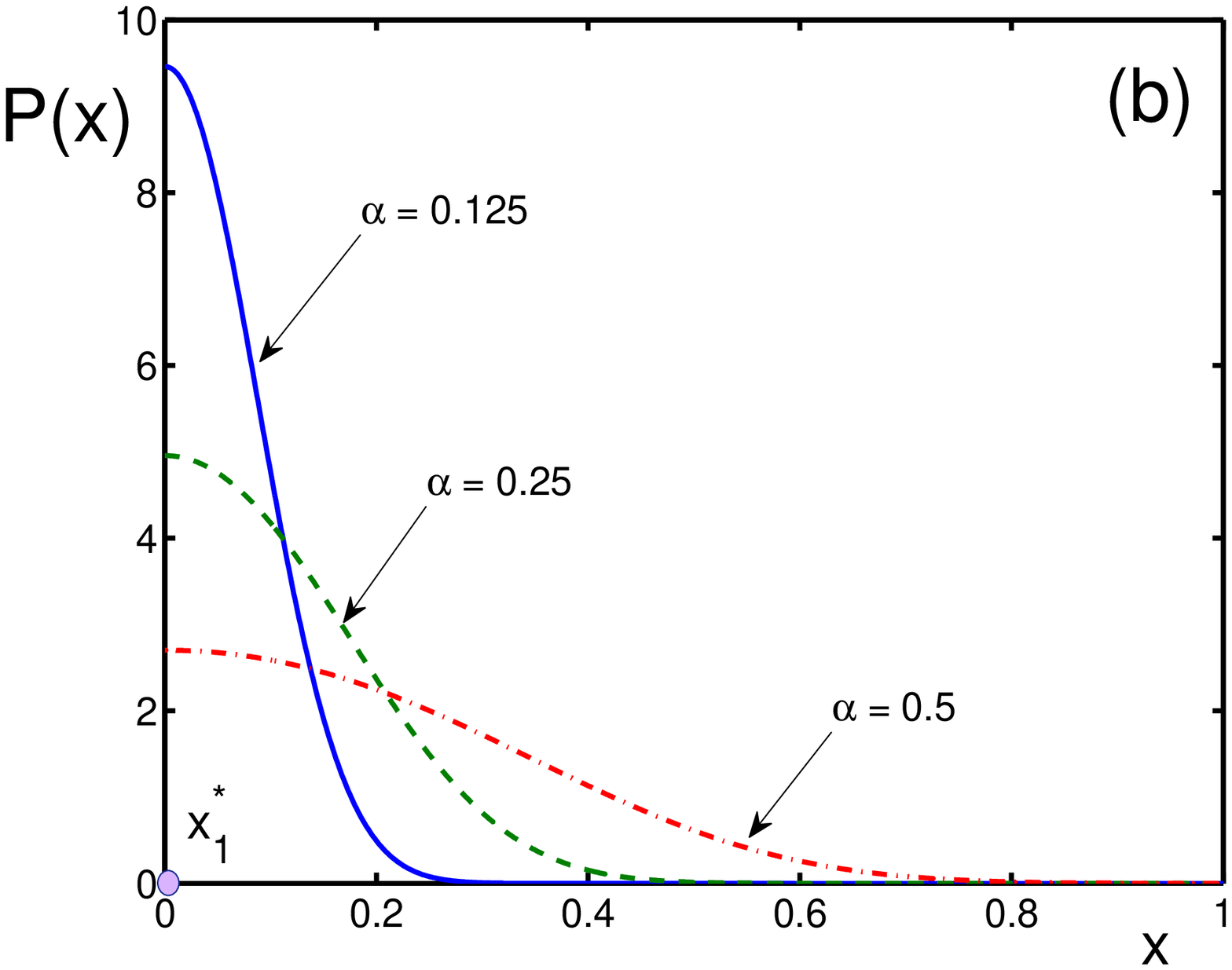} } }
\vspace{9pt}
\centerline{
\hbox{ \includegraphics[width=6.5cm]{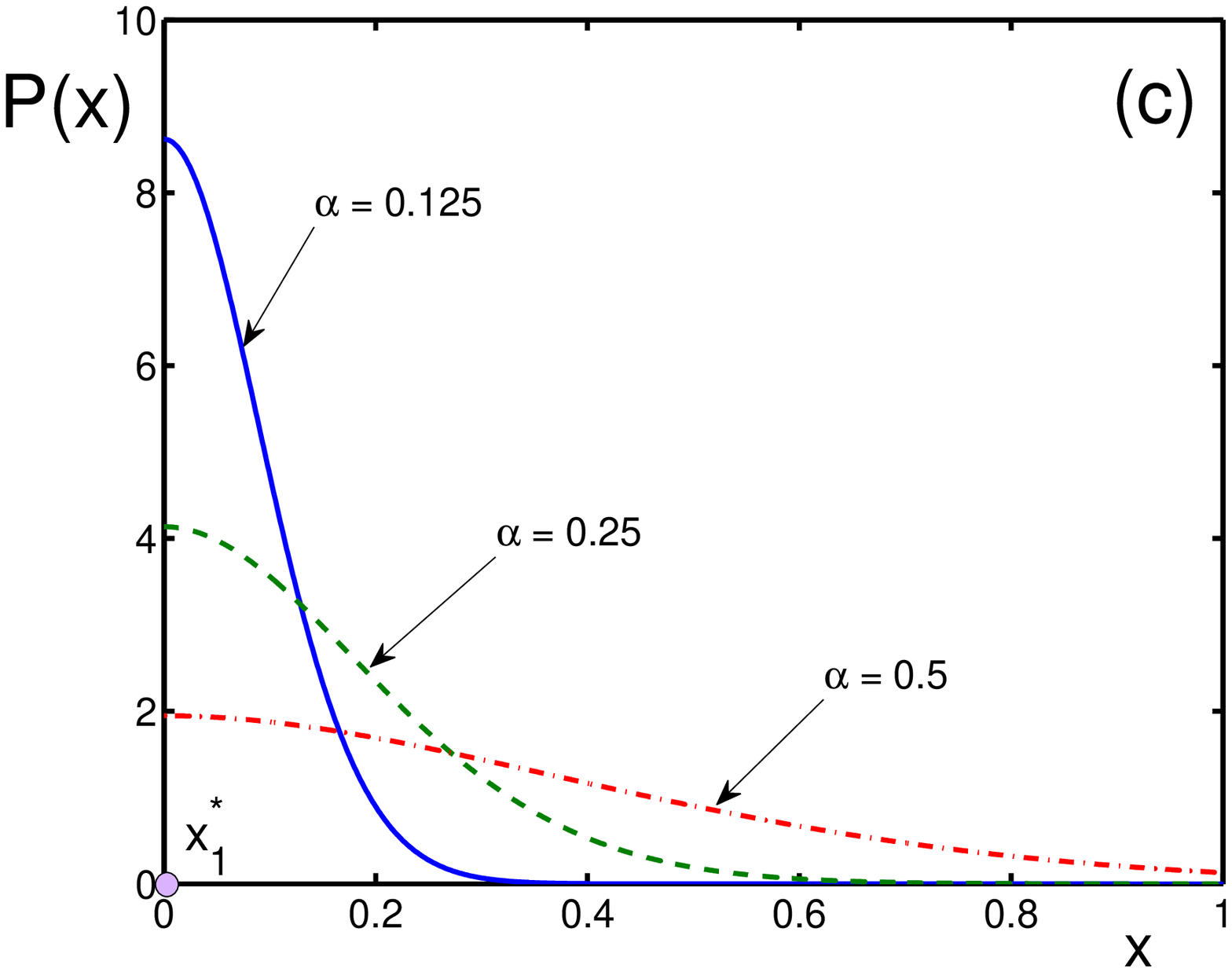} } }
\caption{Distribution $P(x)$, as a function of $x$, in the case of 
a single fixed  point $x_1^* = 0$, for the noise strengths $\al = 0.125$ 
(solid line), $\al=0.25$ (dashed line), and $\al= 0.5$ (dashed-dotted line), 
for different parameters: (a) $\sigma_1 = -1$, $\sigma_2 = 1$, $b = 1$; 
(b) $\sigma_1 = -1$, $\sigma_2 = 1$, $b = -1$; (c) $\sigma_1 = -1$, 
$\sigma_2 = -1$, $b = 1$.}
\label{fig:Fig.18}
\end{figure}

\begin{figure}[ht]
\vspace{9pt}
\centerline{
\hbox{ \includegraphics[width=6.5cm]{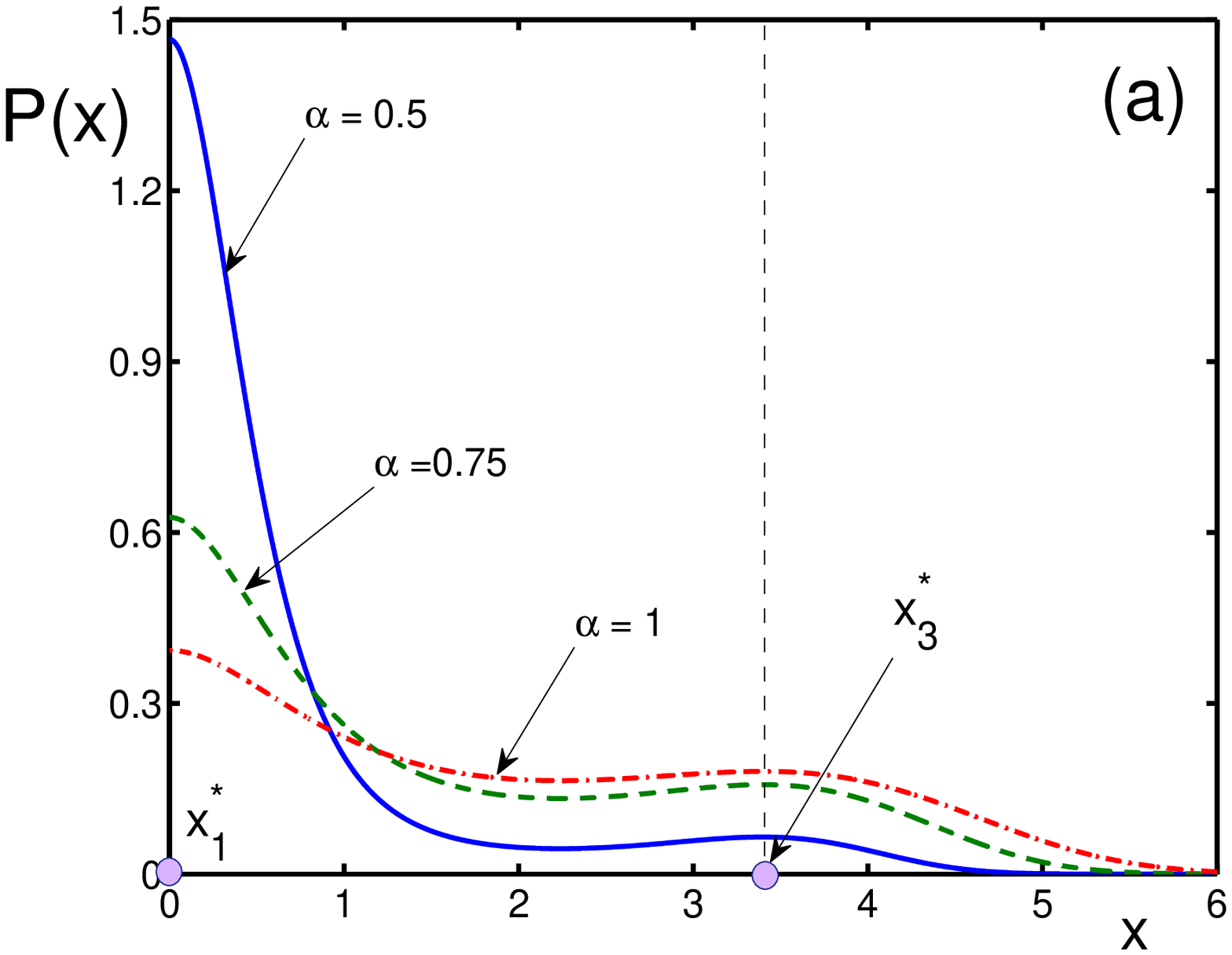} \hspace{2cm}
\includegraphics[width=6.5cm]{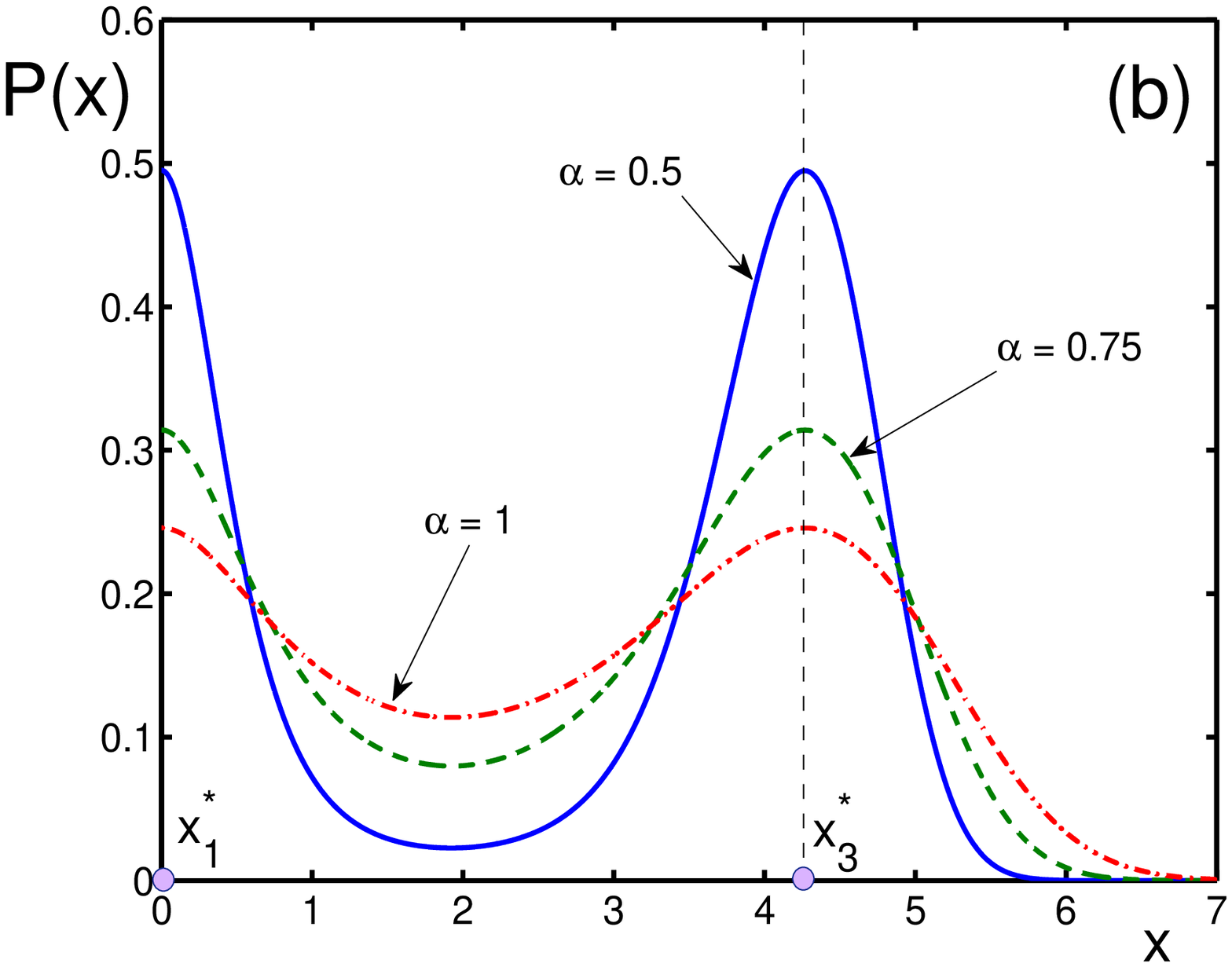} } }
\vspace{9pt}
\centerline{
\hbox{ \includegraphics[width=6.5cm]{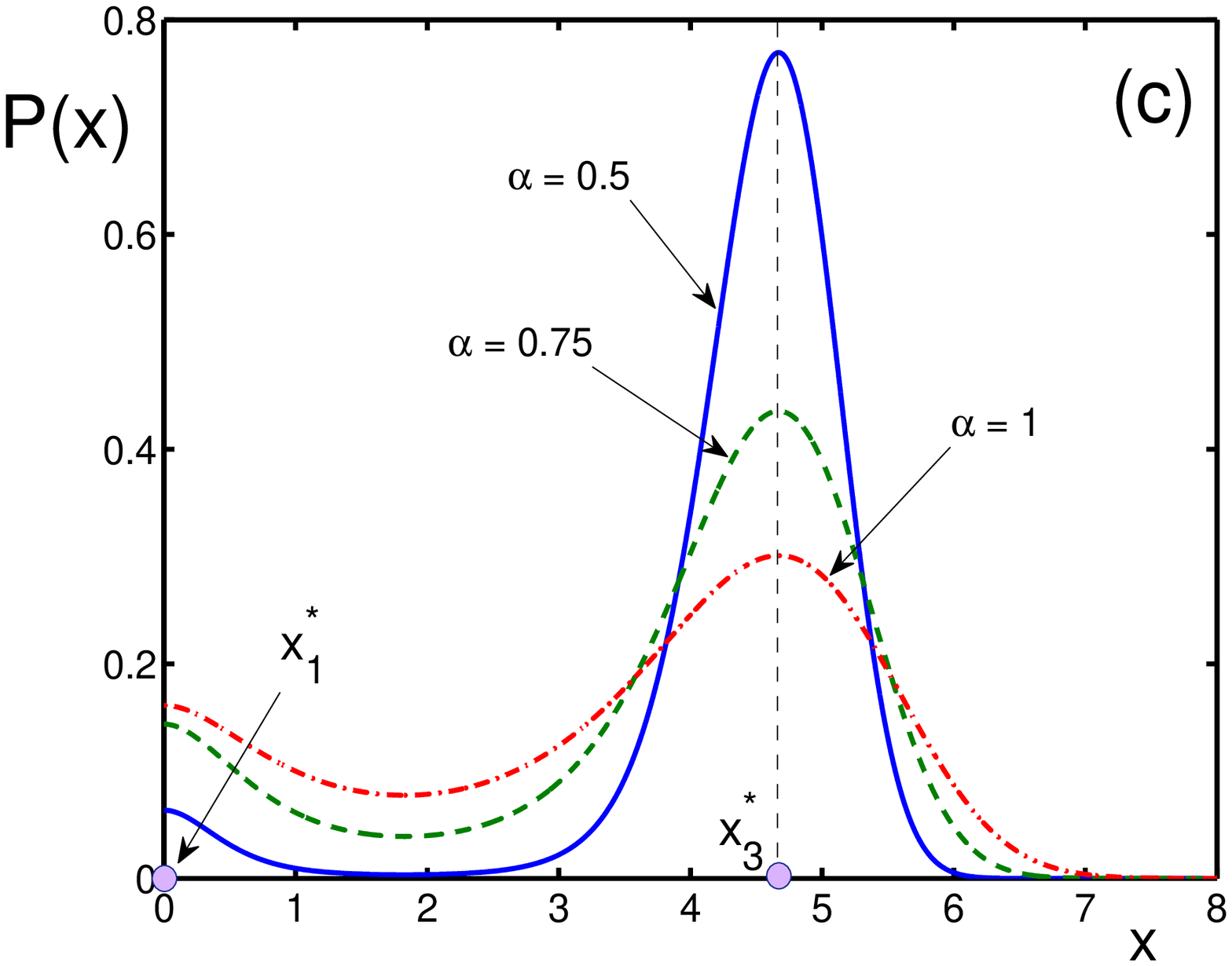} \hspace{2cm}
\includegraphics[width=6.5cm]{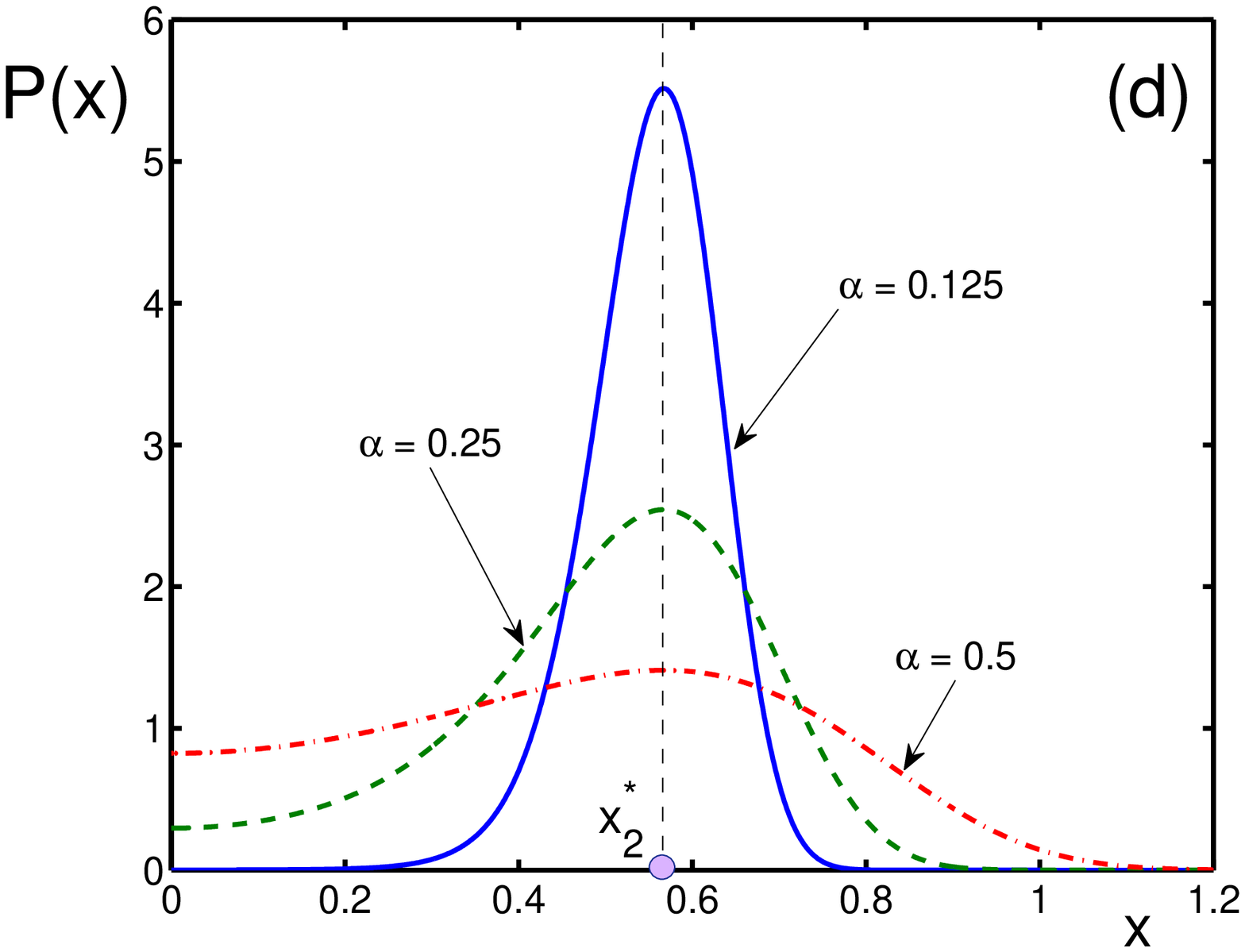} } }
\caption{Distribution $P(x)$, as a function of $x$, in the case of 
either the bistability region with two fixed points $x_1^*=0$ and $x_3^*>0$ 
or for one nontrivial fixed point $x_2^* > 0$. The parameters are: 
(a) $\sigma_1=\sigma_2=-1$, $b=0.36$, with $x_3^*=3.397$, for $\al=0.5$ 
(solid line), $\al=0.75$ (dashed line), and $\al = 1$ (dashed-dotted line);
(b) $\sigma_1 = \sigma_2 = -1$, $b = 0.34$, with $x_3^* = 4.268$, for the 
same noise strengths as in (a); (c) $\sigma_1 = \sigma_2 = -1$, $b = 0.33$, 
with $x_3^* = 4.671$, for the same noise strengths as in (a); 
(d) $\sigma_1 = \sigma_2 = 1$, $b = -1$, with $x_2^* = 0.567$, for the 
noise strengths $\al=0.125$ (solid line), $\al=0.25$ (dashed line), and 
$\alpha = 0.5$ (dashed-dotted line).  
}
\label{fig:Fig.19}
\end{figure}

\section{Prediction of finite-time singularities}

When the solution diverges at a finite time, this is called a {\it finite-time
singularity}. As has been discussed above, as well as in a series of papers
\cite{Kapitza_1,Hern_2,Korotayev_3,Yukalov_28,Yukalov_13,Yukalov_14,Johansen_24,
Fogedby_25,Sornette_26,Sornette_30,Andersen_27,Andersen_29}, a finite-time 
singularity represents a kind of a critical point, where the system experiences 
a transformation to another dynamic state, similarly to the occurrence of phase 
transitions in statistical systems \cite{Yukalov_32,Sornette_33}. Such critical 
points for complex systems, described by the evolution equations of type (7), 
depending on applications, can correspond to the points of overpopulation, firm 
ruin, market crash towards the end of a bubble, earthquakes, and so on. It is 
evident that the possibility of predicting such disasters would be of great 
importance. Here, for the case of finite-time singularities, we show that indeed 
such a prediction can be feasible.

Suppose, we make observations for the behavior of the function of interest $x(t)$,
starting from an initial condition $x_0$, during a short period of time, when this
function can be well characterized by the simple asymptotic form
\be
\label{49}
x(t) \simeq x_0 + c_1 t + c_2 t^2 \qquad (t\ra 0 ) \;   .
\ee
In real life, the coefficients $c_i$ can be defined from the observed data. And 
for the considered equation, they are found by substituting expression (\ref{49}) 
into the evolution equation. Thus for the case $\tau > t_c$, we get
$$
c_1 = \sgm_1 x_0 - \sgm_2 x_0^2 \exp (-bx_0 ) \; ,
$$
$$
c_2 = \frac{c_1}{2}\;
\left [ \sgm_1 - \sgm_2 (2 - bx_0 ) x_0 \exp(-bx_0) \right ] \;    .
$$

If $c_1$ is negative, this means that, in the vicinity of the initial condition 
$x_0$, the function $x(t)$ decreases, hence in the near future, we do not expect 
the occurrence of a singularity, where $x(t)$ would quickly rise. If $c_1$ is 
positive, then $x(t)$ increases, and the singularity is not excluded. To understand 
whether it really happens, we need to extrapolate the asymptotic series (\ref{49}) 
to longer times. A powerful method for extrapolating asymptotic series has been 
developed in Refs. \cite{Yukalov_34,Gluzman_35,Yukalov_36}, being termed the
{\it method of self-similar factor approximants}. This method has been proved to 
be accurate for predicting critical points of different nature, including the
critical points for dynamical systems
\cite{Yukalov_34,Gluzman_35,Yukalov_36,Yukalov_37,Yukalova_38}.

In the framework of self-similar factor approximants, the second-order factor
approximant reads
\be
\label{50}
 x^*(t) = x_0 ( 1 + A t)^n \;   .
\ee
The parameters $A$ and $n$ are defined by expanding (\ref{50}) in powers of $t$ 
and comparing the expansion with the asymptotic form (\ref{49}), which yields
$$
A = \sgm_2 x_0 (1 - bx_0) \exp (-bx_0) \; , \qquad
n = \frac{\frac{\sgm_1}{\sgm_2} - x_0\exp(-bx_0)}{x_0(1-bx_0)\exp(-bx_0) }\; .
$$

Let us recall that, as the analysis of the previous sections shows, for the 
considered case of $\tau > t_c$, the finite-time singularity happens under one 
of the following conditions: either for $\sigma_1 = 1, \sigma_2 = -1, b < 0$ and 
for any history $x_0$, or for $\sigma_1 = \sigma_2 = -1, b < 0$ and $x_0 > x_2^* < e$. 
In both these cases, we find that $A < 0$ and $n < 0$, which allows us to 
rewrite Eq. (50) as
\be
\label{51}
x^*(t) = \frac{x_0}{(1-| A | t)^{| n |} }\;   .
\ee
The latter expression shows that the point of singularity is given by
\be
\label{52}
 t_c^{app} = \frac{1}{| A |} =
\frac{\exp(bx_0)}{\sgm_2x_0(1-bx_0)} \; .
\ee

We have investigated the behavior of formula (\ref{52}) for the different
situations studied in the previous sections. We find that, when the evolution
equation gives a solution $x(t)$ diverging at a finite time, then the predicted
value (\ref{52}) does approximate the real divergence point $t_c$. When the 
solution $x(t)$ is bounded, approaching a stationary state, then either $A$ 
or $n$ is positive, so that the factor approximant (\ref{50}) does not predict 
singularities. And, if the solution $x(t)$ tends to infinity for $t \ra \infty$, 
then the factor approximant (\ref{50}) either does not show a finite-time 
singularity or, in some cases exhibits its appearance. Such artificial 
singularities can be removed by constructing the factor approximants of higher 
orders. In order not to complicate the consideration, here we limit ourselves 
to the second-order factor approximant that does predict the singularity when 
it really happens for $x(t)$.

To estimate the accuracy of the prediction, we calculate the predicted $t_c^{app}$
for different values of the parameters and compare it with the $t_c$ given by the
evolution equation. The results in Fig. 20 demonstrate that the predicted singularity
point $t_c^{app}$ is close to the real $t_c$. The accuracy can be more precisely
characterized by the absolute and relative errors
$$
 \Dlt \equiv |\; t_c^{app} - t_c \; | \; , \qquad
\ep \equiv \frac{\Dlt}{t_c} \times 100\% \;  .
$$
In Table 1, we present such error metrics, with fixed $x_0 = 1$, for varying 
parameters $b$.

\begin{figure}[ht]
\vspace{9pt}
\centerline{
\hbox{ \includegraphics[width=6.5cm]{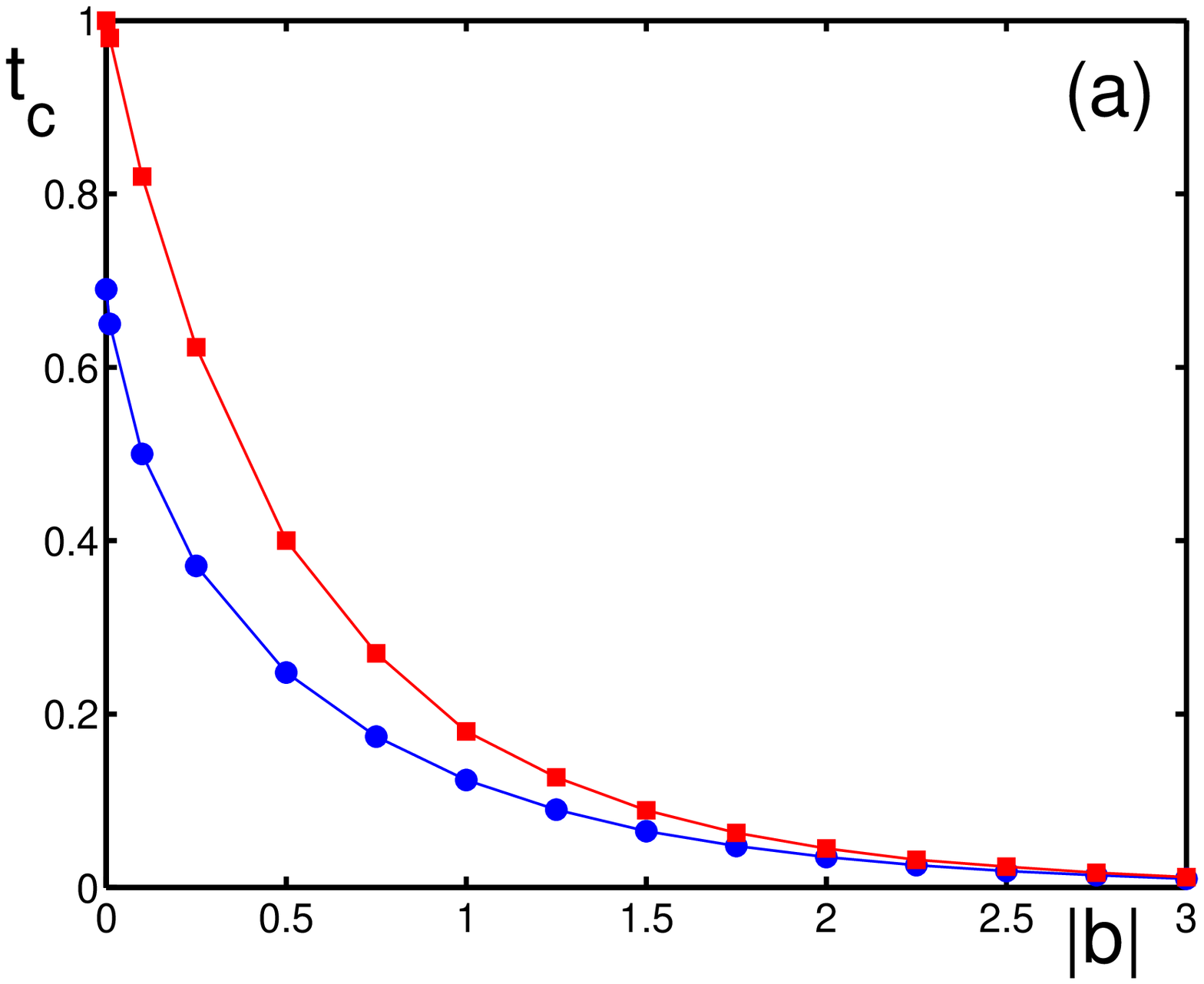} \hspace{2cm}
\includegraphics[width=6.5cm]{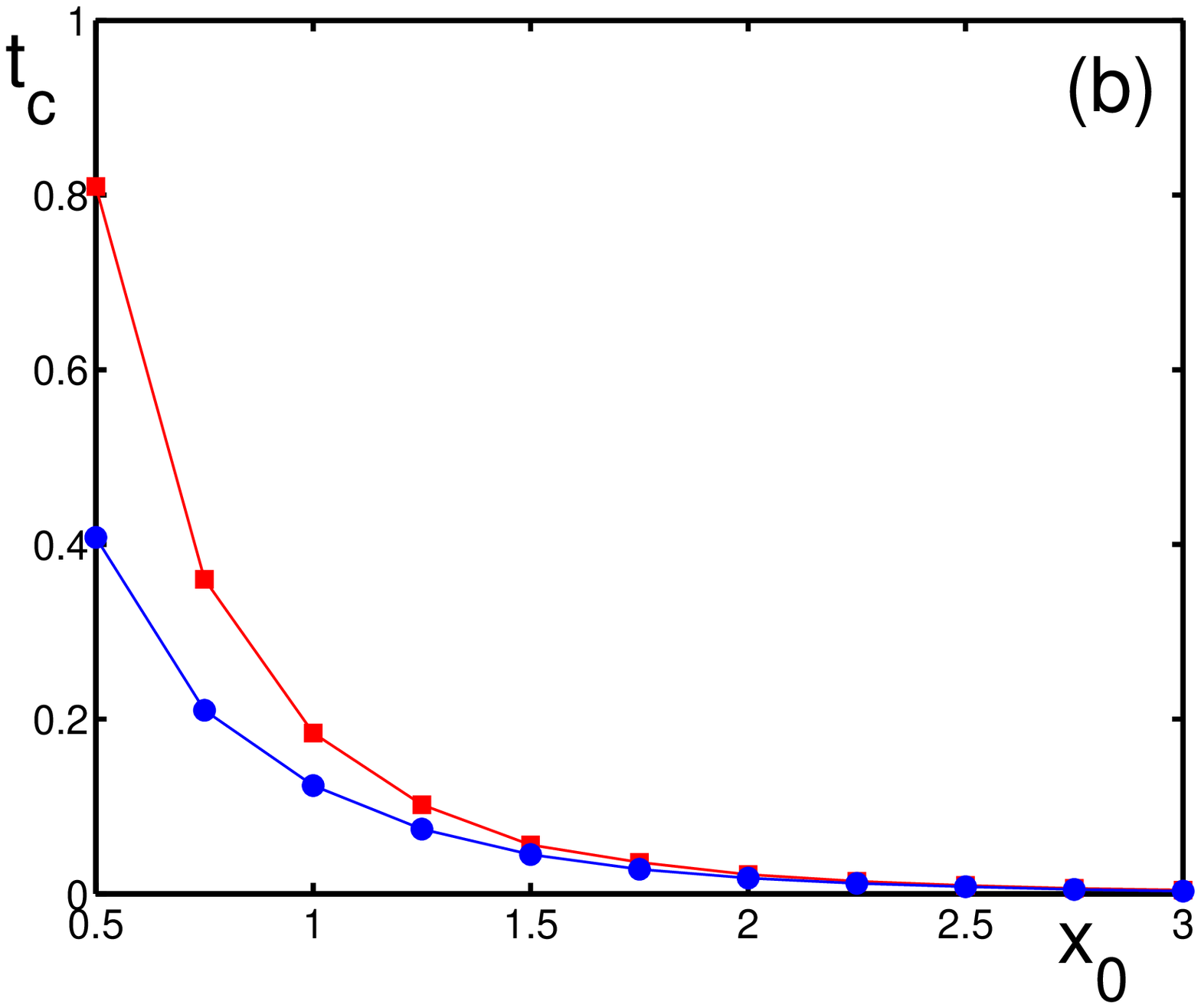} } }
\caption{Predicted singularity times $t_c^{app}$ (red line with squares) 
and real $t_c$ (blue line with circles), as functions of the parameter $b=-|b|$ 
and history $x_0$, under $\sgm_1=1$, $\sigma_2 = -1$ for: (a) varying $|b|$, 
with fixed $x_0 = 1$; (b) varying $x_0$, with fixed $b = -1$.
}
\label{fig:Fig.20}
\end{figure}

\begin{table}[h]
\tbl{The singularity time $t_c^{app}$ predicted by the self-similar factor 
approximant (52), as compared to the exact singularity time $t_c$ following from 
the evolution equation}
{ \begin{tabular}{|l|l|l|c|l|}  \hline
$ b$    & $t_c$  & $t_c^{app}$ & $\ep\%$ & $\Dlt$ \\ \hline
-0.0001 & 0.692  & 1.00       & 45\%    & 0.308 \\ \hline
-0.01   & 0.654  & 0.980      & 50\%    & 0.326 \\ \hline
-0.1    & 0.501  & 0.823      & 64\%    & 0.322  \\ \hline
-0.5    & 0.249  & 0.404      & 62\%    & 0.155  \\ \hline
-1.0    & 0.124  & 0.184      & 48\%    & 0.060  \\ \hline
-1.5    & 0.065  & 0.089      & 37\%    & 0.024  \\ \hline
-2.0    & 0.035  & 0.045      & 29\%    & 0.010  \\ \hline
-2.5    & 0.019  & 0.024      & 26\%    & 0.005   \\ \hline
-3.0    & 0.010  & 0.012      & 20\%    & 0.0020  \\ \hline
-3.5    & 0.0057 & 0.0067     & 18\%    & 0.0010  \\ \hline
-4.0    & 0.0032 & 0.0037     & 14\%    & 0.0005  \\ \hline
\end{tabular} }
\end{table}

Note that $t_c^{app}$ is systematically larger than the true singularity 
time $t_c$, as can be expected from the fact that the sole information used 
in the prediction is the quadratic asymptotic representation (\ref{49}), which 
necessarily underestimates the full strength of the nonlinear feedback leading 
to the singularity. Taking third and higher order terms into account would 
lead to significant improvement of the prediction accuracy. But we believe that 
using the quadratic asymptotic form (\ref{49}) is a realistic proxy for the 
capture of early time dynamics in real life situations. While the relative 
errors are significant (from 14\% to 64\% in the investigated cases), we believe 
that these predictions are useful to provide an approximate estimation of the 
critical time of the singularity. Taking into account more terms in the 
asymptotic expansion for $x(t)$ would improve the accuracy of the prediction,
however involving more complicated expressions for the critical time.

\section{Conclusion}

We have considered the evolution equation describing the population dynamics 
with functional delayed carrying capacity. The linear delayed carrying capacity, 
advanced earlier by the authors, has been generalized to the case of a nonlinear 
delayed carrying capacity. This allowed us to treat the delayed feedback of 
the evolving population on the capacity of their surrounding, by either creating 
additional means for survival or destroying the available resources, when the 
feedback can be of arbitrary strength. This is contrary to the linear approximation 
for the capacity, which assumes weak feedback. The nonlinearity essentially 
changes the behavior of solutions to the evolution equation, as compared to the 
linear case.

The justification for the exponential form of the nonlinearity is based on the
derivation of an effective limit of expansion (5) for the carrying capacity by
invoking the self-similar approximation theory. 

All admissible dynamical regimes have been analyzed, which happen to be of the 
following types: punctuated unbounded growth, punctuated increase or punctuated 
degradation to a stationary state, convergence to a stationary state with sharp 
reversals of plateaus, oscillatory attenuation, everlasting fluctuations, 
everlasting up-down plateau reversals, and divergence in finite time. The theorem 
has been proved that, for the case of gain and competition, the solutions are 
always bounded, when the feedback is destructive.

We have studied the influence of additive noise in two cases: (i) on the solutions 
exhibiting finite-time singularities and (ii) in the presence of stationary solutions. 
For the former case, we found that even a small noise level profoundly affects 
the position of the finite-time singularities. For the later case, we have used 
the Fokker-Planck equation and derived the general condition for the existence of 
a stationary distribution function.

Finally, we showed that the knowledge of a simple quadratic asymptotic behavior 
of the early time dynamics of a solution exhibiting a finite-time singularity 
provides already sufficient information to predict the existence of a critical 
time, where the solution diverges. 

It is necessary to stress that taking into account the nonlinear delayed 
carrying capacity not merely changes quantitatively the behavior of the solutions 
to the evolution equation, but also removes artificial finite-time divergence 
and finite-time death that exist in the equation with the linear form of the 
carrying capacity. For example, the linear carrying capacity can lead to the 
appearance of finite-time singularity or finite-time death even in the case of 
prevailing competition $(\sigma_2 = 1)$, as is found in
\cite{Yukalov_13,Yukalov_14}. But with the nonlinear carrying capacity, as used in 
the present paper, these finite-time critical phenomena are excluded. Now, 
finite-time singularity can occur only in the logically clear case of cooperation
$(\sigma_2 = -1)$. The reason why the linear approximation for the carrying 
capacity leads to such artificial singularities and deaths has been explained in
Sec. 2.1 of the present paper.

\nonumsection{Acknowledgments} 
\noindent 

Financial support from the ETH Competence Center "Coping with Crises in Complex
Socio-Economic Systems" is appreciated.


\end{document}